\documentclass[sigconf]{acmart}

\AtBeginDocument{%
  \providecommand\BibTeX{{%
    \normalfont B\kern-0.5em{\scshape i\kern-0.25em b}\kern-0.8em\TeX}}}

\setcopyright{acmcopyright}
\copyrightyear{2023}
\acmYear{2023}
\acmDOI{XXXXXXX.XXXXXXX}

\acmConference[CHI'25]{The ACM CHI conference on Human Factors in Computing Systems}{April 26--May 1st,
  2024}{Yokohama, Japan}
\acmPrice{15.00}
\acmISBN{978-1-4503-XXXX-X/18/06}

\usepackage{graphicx}
\usepackage{svg}

\usepackage{subcaption}
\usepackage{pdfpages}
\usepackage{hyperref}
\usepackage{lineno}
\usepackage{enumitem}
\usepackage{booktabs}
\usepackage{caption}
\usepackage{array}
\usepackage{enumitem}

\begin{document}
\nolinenumbers

\title[End User Authoring of Personalized Content Classifiers]{End User Authoring of Personalized Content Classifiers: Comparing Example Labeling, Rule Writing, and LLM Prompting}

\author{Leijie Wang}
\email{leijiew@cs.washington.edu}
\affiliation{%
  \institution{University of Washington}
  \city{Seattle}
  \country{United States}
}

\author{Kathryn Yurechko}
\email{kathryn.yurechko@oii.ox.ac.uk}
\affiliation{%
  \institution{University of Oxford}
  \city{Oxford}
  \country{United Kingdom}
}

\author{Pranati Dani}
\email{pdani1@cs.washington.edu}
\affiliation{%
  \institution{University of Washington}
  \city{Seattle}
  \country{United States}
}

\author{Quan Ze Chen}
\email{cqz@cs.washington.edu}
\affiliation{%
  \institution{University of Washington}
  \city{Seattle}
  \country{United States}
}

\author{Amy X. Zhang}
\email{axz@cs.uw.edu}
\affiliation{%
  \institution{University of Washington}
  \city{Seattle}
  \country{United States}
}

\newcommand\deleted[1]{}
\newcommand\leijie[1]{#1}

\renewcommand{\shortauthors}{Wang Leijie et al.}


\begin{abstract}

Existing tools for laypeople to create personal classifiers often assume a motivated user working uninterrupted in a single, lengthy session.
However, users tend to engage with social media casually, with many short sessions on an ongoing, daily basis.
To make creating personal classifiers for content curation easier for such users, tools should support rapid initialization and iterative refinement.
In this work, we compare three strategies---(1) example labeling, (2) rule writing, and (3) large language model (LLM) prompting---for end users to build personal content classifiers.
From an experiment with 37 non-programmers tasked with creating personalized moderation filters, \leijie{we found that participants preferred different initializing strategies in different contexts, despite LLM prompting's better performance. However, all strategies faced challenges with iterative refinement. To overcome challenges in iterating on their prompts, participants even adopted hybrid approaches such as providing examples as in-context examples or writing rule-like prompts.}

\end{abstract}



\maketitle

\section{Introduction}
Today, internet users encounter more content than ever before, ranging from social media posts, blogs, and news articles to chat conversations and emails. To manage this overwhelming and sometimes unwanted information, online platforms provide automated curation systems to categorize, label, and moderate content. Some offer content moderation algorithms to remove harmful posts~\cite{YoutubeAlgorithm, FacebookAlgorithm}, automatic classifiers to organize and prioritize emails~\cite{GmailSpamFilter}, or recommendation algorithms to discover interesting content~\cite{eksombatchai2018pixie}. But these systems are typically centralized and platform-wide, failing to accommodate the diverse preferences of individual users~\cite{neves2018customization,kumar2021designing}.
To support end-user customization, researchers have examined and built a variety of specialized tools that enable users to author personalized content classifiers within social media~\cite{jhaver2022designing, he2023cura, park2019opportunities,jhaver2019human}. They have also explored generic techniques for non-technical people to build their own text classifiers~\cite{hema2006active, cheng2015flock, vsavelka2015applying}.

However, many tools for authoring personal classifiers on social media are designed for users with \textit{heightened motivation}, such as community moderators~\cite{chandrasekharan2019crossmod,jhaver2019human} and high-profile content creators~\cite{jhaver2022designing}. In focusing on these users, such tools neglect the usage patterns of general social media users, who spend significant time on social media over the long term but in many relatively short, fragmented sessions. They also have limited attention and motivation for cognitively demanding tasks~\cite{socialMediaUse, jhaver2022designing}.
As a result, for many users to realistically participate in curation, tools for authoring classifiers should support \textbf{rapid initialization}. As end users oftentimes engage with internet content as a leisure activity, a successful tool should allow them to quickly and intuitively build an initial classifier with decent performance. In contrast, existing systems often demand that users maintain a high level of concentration for extended periods in this process~\cite{kulesza2015principles, stretcu2023agile}. For instance, some treat users as oracles who can continuously provide high-quality input, whereas others overwhelm users with an excessive amount of information~\cite{kulesza2015principles, song2023modsandbox}. 
In addition, curation tools should enable \textbf{easy iteration} to improve initial creations incrementally, as social media users may be more amenable to short tasks spread out over many sessions as opposed to s single lengthy task. Instead, existing systems often assume that users want to create highly performant custom classifiers \textit{in a single sitting}~\cite{kulesza2015principles, vsavelka2015applying, ratner2017snorkel}, for instance, requiring that users carefully debug their classifiers before their deployment. However, social media users naturally audit curation algorithms as they browse their feeds regularly~\cite{shen2021everyday}. Their preferences may also evolve gradually, necessitating continuous iteration~\cite{feng2024mapping}. Thus, a successful tool should enable users to make small adjustments over a period of time and ensure that each change results in the desired incremental improvement.

In this work, we use these system requirements to compare three prominent strategies for creating custom classifiers for content curation: (1) \textit{labeling examples for supervised learning}, (2) \textit{writing and carrying out rules}, and (3) \textit{prompting a large language model (LLM)}. \leijie{In particular, we aim to answer the following two research questions:
\begin{itemize}
    \item Which strategy allows end users to initialize a personal content classifier the most easily?
    \item Which strategy allows end users to iterate on their personal content classifiers after initialization the most easily?
\end{itemize}}

Each of these three strategies has its respective benefits and drawbacks for end-user classifier creation.
In interactive machine learning (IML) research, \textit{labeling examples} is a popular strategy to customize a classifier. It is promising in the social media context due to its simplicity and ability to be easily divided into smaller tasks. However, even though pretrained word-embeddings~\cite{devlin2018bert} and active learning~\cite{settles2009active} reduce the number of labels needed to train a classifier, labeling examples can still be tedious and inefficient~\cite{das2013end, kulesza2014structured}, so this technique is typically not available for social media users to create personal classifiers from scratch.
In contrast, the most common strategy deployed on social platforms today is \textit{writing rules}. Most platforms allow users to curate content through simple keywords~\cite{jhaver2022designing}. AutoMod, the most widely used content moderation tool on Reddit~\cite{jhaver2019human, chandrasekharan2019crossmod}, further enables community moderators to write regular expressions to remove inappropriate posts~\cite{jhaver2019human}. Despite the transparency that rules afford, users often struggle to write rules that account for nuanced contexts and to further refine them~\cite{jhaver2022designing,song2023modsandbox, chandrasekharan2019crossmod}. 
Finally, with recent advances in LLMs, we additionally investigate the technique of customizing zero-shot or few-shot classifiers by \textit{prompting an LLM} in natural language~\cite{kumar2023watch, OpenAI2024GPT4}. While LLMs have the potential to learn nuanced preferences quickly~\cite{wang2019survey}, their effective use may require prompt engineering skills beyond the capabilities of many end users~\cite{sclar2023quantifying, lu2021fantastically}. 

We conducted a within-subjects lab experiment to evaluate these three strategies.
For each strategy, we implemented a content curation system with state-of-the-art features to ensure a fair comparison.
We then invited 37 non-technical social media users to build a personal classifier using each system. 
Each participant first manually labeled 100 comments sourced from YouTube videos on political topics to create a personal ground truth dataset, which allowed us to evaluate the performance of their classifiers. Then, in a randomized order, they used each of the three systems to create personal classifiers for removing unwanted YouTube comments, with a 15-minute time limit per system. 
We logged the actions participants took in each condition and tracked the performance of the classifiers as participants built them. After each condition, participants filled out a survey to report their subjective user experience.
Additionally, we conducted 13 semi-structured interviews with a subset of participants after all conditions were complete to better understand their challenges in communicating initial preferences and iterating on classifiers in different conditions.

\leijie{Through a quantitative and qualitative analysis, we compared how these three strategies support end users to initialize and iterate on their personal content filters. In more detail, we present the following findings:
\begin{itemize}[leftmargin=*]
    \item We found that writing prompts generally allowed participants to create personal content filters with higher performance more quickly, primarily driven by having the highest recall despite comparable precision across conditions.
    \item Nonetheless, participants expressed reason to prefer each of the strategies when initializing a filter:
    \begin{itemize}
    \item When preferences were \textit{ill-defined but intuitive}, labeling examples was considered the easiest method;
    \item When participants had \textit{specific topics or events} they were trying to filter, writing rules was considered easiest; and
    \item When preferences were both relatively \textit{well-defined and general}, writing prompts was considered easiest.
    \end{itemize}
    \item All strategies struggled with iterative refinement. Participants particularly found it challenging aligning LLMs with their nuanced preferences, as recall and precision often plateaued within the first five minutes.
    \item  In their effort to improve their filters while writing prompts, participants often turned to approaches that looked like the other strategies, for instance, by adding misclassified examples directly as in-context examples or writing prompts resembling rules.
    \item Finally, no single strategy was clearly preferred across the board. We catalog participants' needs beyond ease that led to a divergence in preferences, including their varying tolerance for viewing toxic content and prioritization of precision versus recall.
\end{itemize}}


Our findings offer a roadmap for future end-user content curation systems and have broader implications for facilitating more efficient and effective communication between end users and classifiers. 
First, while LLMs are a promising tool for enabling personalized content classifiers, we find that labeling examples requires less cognitive effort from users, and writing rules provides users with more control and transparency over their classifiers. 
Given the diversity of content curation scenarios and user preferences, future content curation systems should provide end users with more flexible strategies to create and iterate on their classifiers.
For instance, while writing prompts could be the default strategy given its superior overall performance, future hybrid systems could also suggest possible prompts based on example labels to help users who have difficulty articulating their intuitions, or allow rule authoring for preferences about specific topics or events.
Hybrid approaches such as these could bring together the benefits of all three strategies and facilitate more effective human-LLM collaboration.

\leijie{
In summary, we made the following two contributions. 
First, this paper explores how existing interactive machine learning strategies fail to address the diverse and dynamic user needs of content curation through a rigorous within-subject experiment.
Second, it provides a nuanced perspective on the use of LLMs in content moderation. This paper highlights their limitations in bootstrapping and iteration and offers design implications to better support effective human-LLM collaboration.
}


\section{Related Work}

\subsection{End User Customization of Content Curation}
Content curation is an important process for internet users who seek to classify desired or undesired content on social media~\cite{lee2019consumptive, jhaver2023personalizing, klug2023algospeak}, filter out harassment or spam from their email folders~\cite{park2019opportunities, mahar2018squadbox}, manage online chats~\cite{volk2024selecting, wang2023reporting}, or find relevant news articles~\cite{volk2024selecting, schmitt2018toomuch}. However, content curation tools are often centralized and platform-wide, leaving limited room for end-user customization~\cite{neves2018customization}. For example, one-size-fits-all moderation algorithms often fail to account for the diverse perceptions of toxicity across countries~\cite{jiang2021understanding} and communities~\cite{weld2022makes}. Similarly, some users want to categorize their emails into more fine-grained folders other than the default ``Spam'' or ``Promotions'' folders~\cite{park2019opportunities, yoo2011modeling}. 
As a result, there have been growing calls to empower end users to customize content curation classifiers~\cite{jhaver2023personalizing, jhaver2023users}. 
In particular, researchers have argued for personal content moderation tools that enable users to customize their social media feeds.
Users may wish to remove unwanted content from their feeds on platforms like Instagram, TikTok, and X/Twitter~\cite{jhaver2023personalizing} or to moderate the comment sections of their YouTube videos or their Twitch streams~\cite{jhaver2022designing}. In both cases, end users act as moderators that remove content according to their preferences.

Researchers and practitioners have experimented with offering custom content curation tools. However, these tools often struggle to navigate the trade-off between flexibility of customization and ease of use.
Some tools offer extensive customization but are too complex for casual users. For example, Reddit’s AutoMod~\cite{jhaver2019human} allows community moderators to write regular expressions in YAML to remove inappropriate posts, but only users with technical knowledge of regular expressions and programming syntax can fully utilize AutoMod's flexibility~\cite{song2023modsandbox}. 
Conversely, other tools are more simplistic in design. Many of them involve pre-trained classifiers for platform-defined or researcher-defined concepts such as racism, misogyny, and political views, with users only able to adjust their sensitivity to each concept~\cite{chandrasekharan2019crossmod, Instagram2024, bhargava2019gobo, Bodyguard2024}. However, this approach fails when users have differing definitions of a concept or have a new concept~\cite{jhaver2023personalizing}.
Recently, research has shown that LLMs can outperform state-of-the-art classifiers in detecting toxic content with natural language prompts~\cite{kumar2023watch, OpenAI2024GPT4}. But LLMs are not exempt from the customization versus ease of use trade-off: when applied toward more complex and nuanced curation preferences that go beyond simple hate speech or toxicity filtering, LLMs sometimes perform even worse than a coin toss~\cite{kumar2023watch}. In this work, we compare common content curation techniques in terms of this trade-off through an experiment using the case of personal content moderation, with the eventual goal of designing a system that offers \textit{both} customization and ease of use for a broad range of end users. 

\subsection{Strategies that Support End Users in Customizing Classifiers for Content Curation}

\deleted{Despite the ubiquitous use of machine learning (ML) algorithms in various domains, only a small group of people with technical expertise possess the skills to develop these algorithms.}
Interactive machine learning (IML) seeks to democratize ML training with humans-in-the-loop, enabling non-experts to participate through ``rapid, focused, and incremental model updates''~\cite{amershi2014power}. IML enables end users to build classifiers in three critical phases ~\cite{amershi2011designing, stumpf2009interacting}: (1) Users examine a model’s outputs across a set of examples; (2) Users then offer feedback to the model, guiding its learning in the desired direction; and (3) Users assess the model's overall performance and decide whether to conclude the training process or to offer the model more feedback.

At the heart of these phases are \textit{teaching vocabularies}: the frameworks through which end users structure their feedback~\cite{porter2013interactive}, including assigning labels, selecting features, or indicating error preferences~\cite{kapoor2010interactive}. Recently, prompts have emerged as another promising way for end users to communicate with LLM-based models. Through the lens of teaching vocabularies, we characterize existing content curation systems by three primary strategies that they leverage to support end-user customization: \textit{labels}, \textit{features or rules}, and \textit{prompts}. In the following, we discuss each strategy and its relationship to content curation.

\subsubsection{\textbf{Labels}}

Many IML systems consider end users to be oracles who provide correct labels. Techniques like pre-trained word embeddings~\cite{devlin2018bert} and active learning~\cite{settles2009active} are often used to minimize the number of labels required to effectively train a ML algorithm~\cite{settles2009active}. An active learning algorithm iteratively picks examples for labeling from an unlabeled dataset through query sampling strategies, such as uncertainty sampling~\cite{lewis1995sequential} and query-by-committee~\cite{seung1992query}. Previous studies have shown that people find labeling effective for communicating with opaque, ``black-box’’ algorithms~\cite{jhaver2023personalizing, feng2024mapping}. However, researchers have noted that end users might find the repetitive labeling of data tedious and non-transparent~\cite{das2013end}. There is also a risk of users applying labels inconsistently due to their evolving preferences~\cite{kulesza2014structured}.

\subsubsection{\textbf{Features or Rules}}

In response to the aforementioned criticisms of the labeling approach, a body of research focuses on feature-level human input~\cite{hema2006active, wallace2012deploying, vsavelka2015applying,cheng2015flock}. Here, the term \textit{``feature''} refers to an attribute of a data instance that ML algorithms use to predict its class label. In comparison, \textit{``rules''} are often logical statements that directly map to a class label without relying on ML algorithms. In systems that support technical users to build a text classifier via feature-level input, features and rules often take the same form: the mentions of keywords. 
By incorporating features or rules into different algorithm architectures, these systems can offer varying degrees of transparency.
Rules can offer the highest transparency by directly classifying text without using any ML models. 
Features, on the other hand, can be incorporated into transparent ML algorithms, such as Decision Trees, Naive Bayes algorithms~\cite{kulesza2015principles}, and Support Vector Machines~\cite{vsavelka2015applying}. 
In addition, because end users might produce noisy and inconsistent features~\cite{das2013end, boecking2020interactive}, user-generated features can also be used to label data. Weak supervision algorithms then learn from these ``soft labels’’~\cite{ratner2017snorkel}. However, these indirect approaches sacrifice the transparency and control that users often assume with feature-level input. For this reason, many systems designed for end users often integrate features with transparent algorithms to enhance users' mental models of such systems~\cite{kulesza2015principles, vsavelka2015applying}.
Although empirical results regarding whether feature-level input from end users improves performance have been mixed~\cite{hema2006active, das2013end, wu2019local}, researchers generally agree that feature-level input requires fewer user actions to achieve comparable results to example labeling. However, features can be seen as too granular and therefore not generalizable enough, posing a challenge for end users in creating effective classifiers~\cite{Chandrasekharan2019, song2023modsandbox}.

\subsubsection{\textbf{Prompts}}

LLMs have exploded in popularity due to their proficiency in a broad range of natural language processing tasks, such as sentiment analysis and machine translation. Compared to features, natural language prompts can convey richer user guidance to models and thus promote greater generalizability across various contexts~\cite{sahoo2024systematic}. Rather than having to train a new model for every custom task, users can simply customize LLMs by feeding them prompts at run time. Such an ability to recognize the desired task on-the-fly is called \textit{in-context learning} (ICL)~\cite{brown2020language}.
Developing effective prompts is crucial to leveraging LLMs in content curation~\cite{min2022rethinking, mishra2021cross}. To date, the most common patterns for prompting are \textit{zero-shot} or \textit{few-shot} prompts. Zero-shot prompts give instructions for a task without any specific examples for training~\cite{wang2019survey}. Research indicates that the performance of zero-shot prompts can be enhanced by iteratively refining task instructions~\cite{efrat2020turking} or breaking down tasks into simpler subtasks~\cite{wei2022chain}. Alternatively, few-shot prompts incorporate a few input-output examples to showcase the desired pattern to which LLMs should adhere~\cite{sahoo2024systematic, brown2020language}. The quality of few-shot prompts relies heavily on the selected examples~\cite{rubin2021learning}. \deleted{While few-shot prompts generally outperform zero-shot prompts, the simplicity of composing natural language instructions without the need for selecting examples makes zero-shot prompting an attractive option for end users.}

\leijie{Despite the apparent simplicity of prompting, prior research has documented various challenges that impede end users in creating and refining their prompts. For example, when designing prompts for robust chatbots, users could often achieve 80\% of their design goals but faced significant difficulty addressing remaining issues~\cite{zamfirescu2023herding}. Additionally, end-users often designed prompts opportunistically and thus struggled to make robust, systematic progress~\cite{zamfirescu2023johnny}. Our research extends this line of work by investigating how end-users designed prompts in the setting of personal content curation. Specifically, we explore how approaches like writing rules or labeling examples can help users more easily and systematically bootstrap and iterate on their prompts.} 

\subsection{\textbf{How Existing IML Systems Overlook End User Needs in Content Curation}}

\leijie{Researchers in End-User Programming and IML have documented various challenges end users face when building algorithms. For instance, users might struggle to find effective ways to teach the algorithm (i.e., use barrier) or diagnose why the algorithm fails (i.e., understanding barriers)~\cite{ko2004six}. As a result, researchers have advocated for supporting non-experts through more intuitive approaches to model building\cite{yang2018grounding}. Building on this line of research, our work situates these challenges within the dynamic and evolving needs of content curation. Specifically, we argue that existing IML systems often make assumptions about end users that do not translate well into the context of content curation.}

First, these systems often require end users to spend hours of dedicated time creating their custom classifiers~\cite{kulesza2015principles, stretcu2023agile}. However, users typically engage with social media as a leisure activity~\cite{feng2024mapping, jhaver2022designing}. 
For example, a recent survey notes that the top two reasons why people use social media are to keep in touch with friends and to fill spare time~\cite{socialMediaUse}. Research also suggests that internet users often rely on default settings, despite acknowledging the benefits of customization controls~\cite{shah2008software, vaccaro2018illusion}. Ultimately, users are only willing to invest effort in creating custom classifiers if there are proportional improvements to their content feeds~\cite{jhaver2023personalizing}.

Additionally, existing IML systems require users to be highly dedicated to creating their custom classifiers. 
For instance, some systems treat users as infallible experts who can continuously provide high-quality labels or rules~\cite{kulesza2014structured}, while others present an overwhelming amount of information intended to facilitate the teaching process~\cite{kulesza2015principles, song2023modsandbox}. 
Even though weak supervision algorithms can learn from noisy user annotations, they often require more annotations to compensate for their low quality~\cite{boecking2020interactive}. Such an expectation of high commitment could thus intimidate users\textemdash who are likely to be casually using classifiers~\cite{feng2024mapping, jhaver2022designing}, accessing them on mobile devices, or using them in between other activities and with distractions~\cite{carstens2018social}\textemdash from using these systems at all.

Finally, IML systems often assume that users want to create highly performant classifiers at their inception~\cite{kulesza2015principles, vsavelka2015applying, ratner2017snorkel}, but users of content curation tools prefer to iteratively refine their classifiers~\cite{song2023modsandbox, chandrasekharan2019crossmod}. 
For example, instead of carefully considering all possible mistakes when initially creating their classifiers, social media users naturally audit their deployed classifiers as they browse content feeds for entertainment~\cite{shen2021everyday}. Members of online communities also tend to help community moderators collect mistakes via user reports~\cite{crawford2016flag}. 
This on-the-go monitoring signals the need for content curation classifiers to adapt to constant distribution shifts and fluid curation preferences: a new political event or video theme could spark community discussions that were unobserved or nonexistent when a user initially developed their classifier~\cite{zhong2020capturing}. Users' curation preferences may also evolve over time, sometimes even depending on their moods~\cite{feng2024mapping}. However, iterating a classifier remains challenging in IML research~\cite{piorkowski2023aimee}. Traditional ML algorithms often require users to provide sufficient new training data or relabel existing training data to reflect updated information. While several systems enable users to update a model via rules~\cite{piorkowski2023aimee} or natural language descriptions~\cite{lee2024clarify} by adjusting the weights of training data, it remains to be explored how these methods could be effectively applied in the context of content curation.

\section{Methods}
Drawing inspiration from research in interactive machine learning and in-context learning, we identified three primary strategies, or \textit{``teaching vocabularies,''} through which users can convey their personal preferences to algorithms: \textit{labeling examples}, \textit{authoring rules}, and \textit{writing prompts}. 
Each teaching vocabulary can be leveraged most effectively by a corresponding backend architecture: black-box ML algorithms, transparent ML algorithms, and LLMs respectively. From the perspective of users, each teaching vocabulary provides different levels of control and requires varying degrees of effort to learn and master. 
Corresponding to these three teaching vocabularies, we developed three representative systems enabling end users to create binary text classifiers, using the scenario of personalized content moderation for YouTube comments. Each system is either a popular or potentially popular tool for end users to customize moderation algorithms.

We conducted a within-subjects experiment with 37 participants with no programming experience to comparatively evaluate these three personal moderation systems for everyday users. Participants used each of the systems\textemdash in a randomized order\textemdash to build a classifier within a fixed amount of time, resulting in three experimental conditions. Throughout the duration of each condition, we collected the performance of participants' classifiers every half-minute in order to understand how well each system enables rapid initialization and iterative improvement over time.
After each condition, we collected participants' self-reported usability ratings of the system in that condition to understand their experiences and challenges that they encountered. For 13 participants, we additionally conducted a semi-structured interview after the experiment was over to understand their challenges in creating and iterating personal classifiers.
We further describe our system implementations and experiment design below.

\begin{figure*}
    \centering
    \includegraphics[width=.9\textwidth]{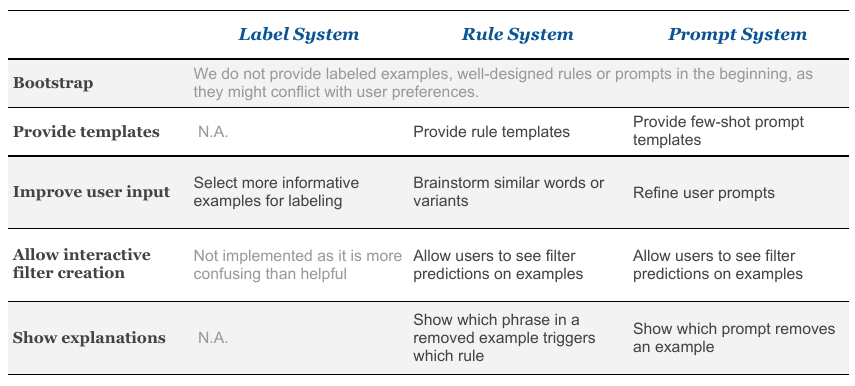}
    \caption{\textbf{Comparison of Implemented Features in Each System}. Light gray texts indicate that we decided not to include the feature in the corresponding system. }
    \hfill
    \label{system features}
    \
\end{figure*}

\subsection{Experiment Systems}
Through pilot studies (detailed in Appendix A), we iteratively implemented three systems corresponding to the three primary teaching vocabularies: \textsc{Label System} (labeling examples to train a supervised model), \textsc{Rule System} (writing executable rules), and \textsc{Prompt System} (writing prompts for an LLM). To ensure the validity of our experiment, we established the following criteria in our system implementations.

First, we \textit{restricted the form of user input} in each system to its designated teaching vocabulary. 
While we acknowledge that an ideal personal moderation system might accept multiple forms of user input, we note that allowing varied types of user input in a single system could obscure the comparison between the three teaching vocabularies in our analysis.
For instance, in a label-based system, allowing users to influence algorithm predictions by highlighting keywords in examples could blur the lines between teaching through labels and teaching through rules (since keywords act as rule indicators).
Similarly, if users are allowed to instruct an LLM to enumerate all possible offensive words in a rule-based system, then users would essentially use both rules and prompts to interact with the algorithm.

Second, we ensured that \textit{similar functionalities} were integrated across all systems so that no condition had an unfair advantage, provided that such functionalities did not compromise the user experience. For instance, both the \textsc{Rule System} and the \textsc{Prompt System} enable users to interactively view how their systems label examples from the training dataset. However, we did not include this feature in the \textsc{Label System} because our pilot studies indicated that displaying system predictions on examples during the active learning process tended to confuse rather than assist end users.

Finally, we equipped each system with \textit{state-of-the-art features} to assist end users in creating their moderation classifiers. For example, the \textsc{Rule System} suggests synonyms for phrases that users have already input, and the \textsc{Prompt System} helps rephrase users' prompts. 
We chose to implement features that involve relatively trivial and already established improvements to the basic authoring interaction to improve performance or usability~\cite{jhaver2022designing}.
Note that this approach does not conflict with our first criterion, as we only enhance user input and do not permit multiple forms of user input.
Thus, our three systems present our attempt to provide the best versions of each strategy while keeping each strategy distinct and functionalities similar for a fair comparison. 
In Figure~\ref{system features}, we summarize the features we chose to implement for each system.
In the following, we discuss the detailed implementation of each system, highlighting how the systems are optimized to support users in the context of personal moderation.

\subsubsection{\textbf{\textsc{Label System}: Train a ``Black-box'' ML Algorithm by Labeling Examples.}}
\begin{figure*}
    \centering
    \includegraphics[width=.7\textwidth]{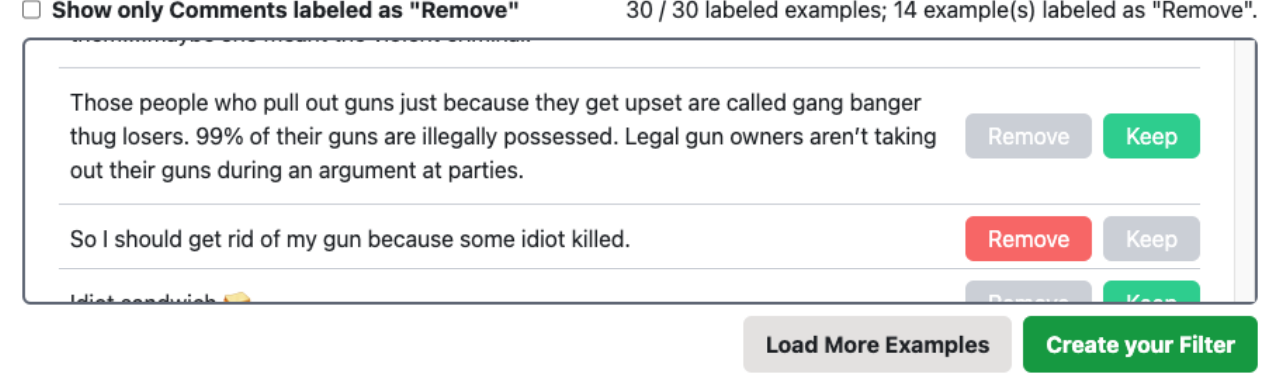}
    \caption{\textbf{The \textsc{Label System}}. Users label each example as  ``Remove'' or ``Keep'' and then load more examples, which are selected via an active learning approach. Since users might refine their criteria during the labeling process~\cite{kulesza2014structured}, they can review examples that they have labeled as ``Remove'' to ensure the consistency of their labels.}
    \hfill
    \label{example labeling}
\end{figure*}

With the \textsc{Label System}, users label example comments to create a personal moderation classifier. We experimented with various models using the Jigsaw toxicity dataset~\cite{borkan2019nuanced} and opted for a combination of Sentence Transformer + Naive Bayes, which had the highest overall performance.
We incorporated active learning to enhance our system, specifically employing uncertainty sampling~\cite{settles2009active}. After a user labels a set of examples, we train the algorithm on these labels and calculate the predicted positive probabilities for the remaining examples in the training dataset. We subsequently sample examples with the highest label uncertainty. In the context of Naive Bayes, examples with prediction scores near 0.5 are deemed the most uncertain. The frontend web interface, depicted in Figure \ref{example labeling}, allows users to label examples selected by active learning and then train the backend algorithms.

\begin{figure*}
    \centering
    \includegraphics[width=\textwidth]{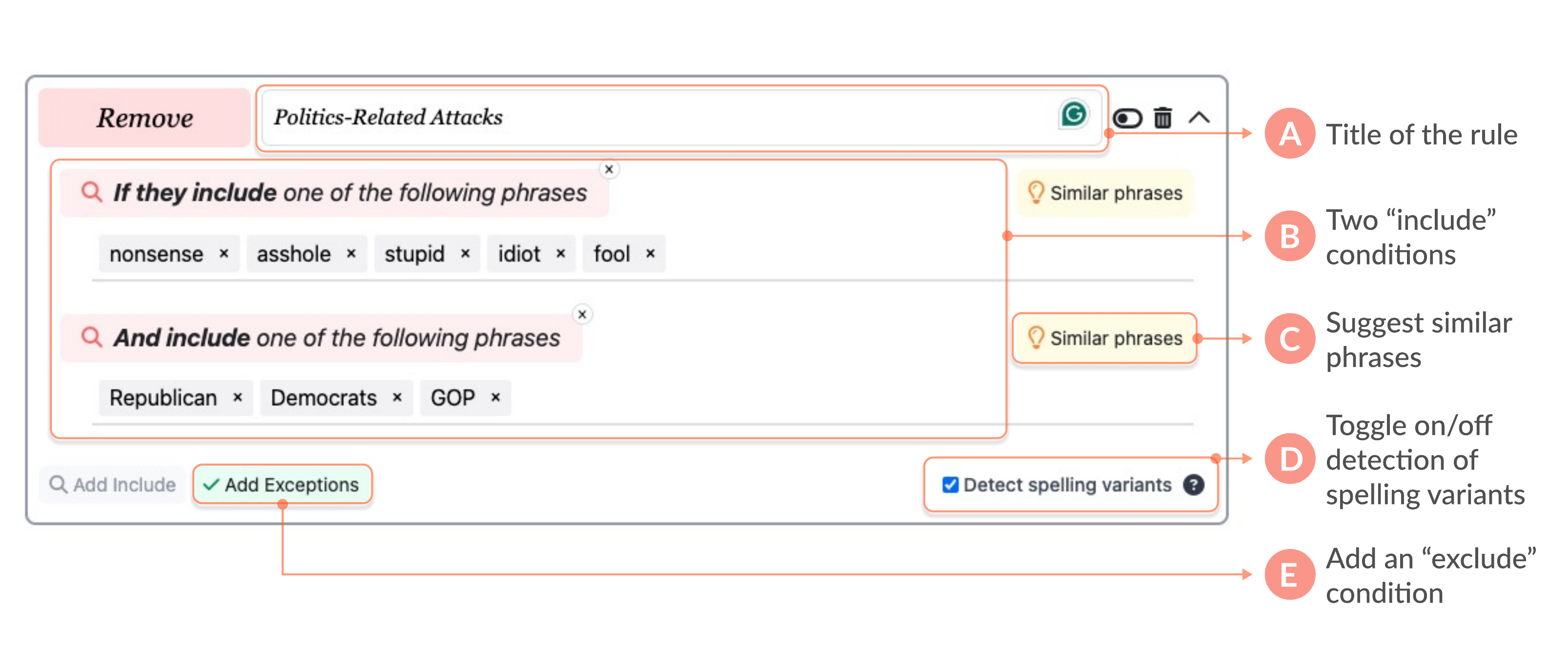}
    \caption{\textbf{An Example of a Rule in the \textsc{Rule System}}. Suppose a user wants to remove texts about politics-related attacks. They can set up an ``include'' condition for attacks and then another ``include'' condition for mentions of politics, which are combined with a logical AND. The user can also add an ``exclude'' condition to specify an exception, such as keeping texts about gun rights even if those texts satisfy their ``include'' conditions.
    The ``similar phrases'' feature leverages an LLM to suggest similar phrases to the existing ones. Detecting spelling variants catches phrases that repeat letters, replace characters with similar characters, contain different forms of a noun or verb, etc.}
    \hfill
    \label{rule block}
\end{figure*}

\subsubsection{\textbf{\textsc{Rule System}: Create a Transparent Algorithm by Authoring Rules}}

\begin{figure*}
    \centering
    \includegraphics[width=.85\textwidth]{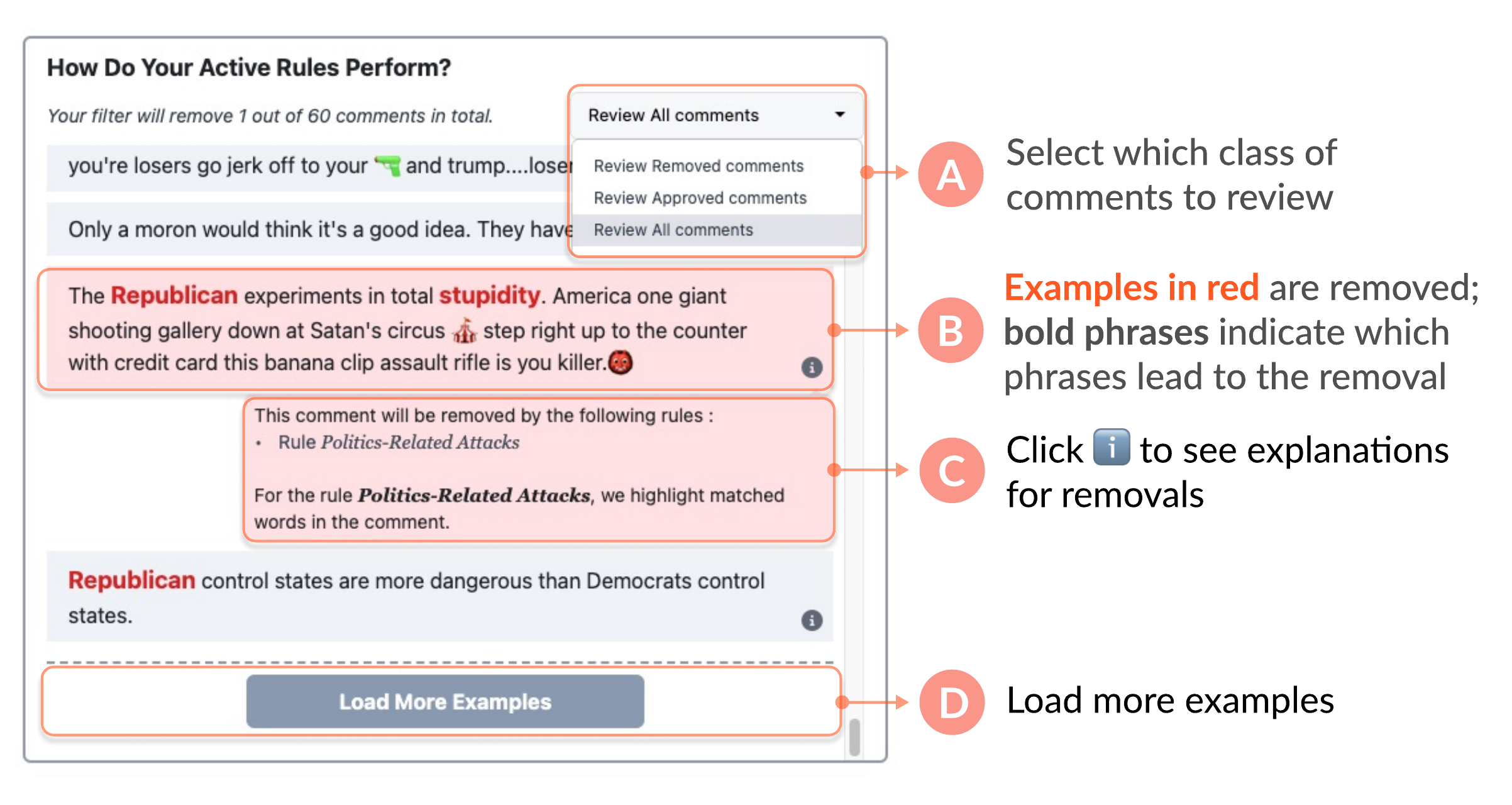}
    \caption{\textbf{The Example Section in the \textsc{Rule System} or \textsc{Prompt System}}. Users can review how their rules/prompts performed on the training dataset. They can review all examples or only removed or approved ones. For each removed example, the system explains which rule/prompt removed it. In addition, the \textsc{Rule System} further explains which phrase in the example triggered the removal. For instance, this figure indicates that the example in red was removed by the rule ``Politics-Related Attacks'' because it includes two phrases that the user associates with personal attacks and political parties respectively. The example below the red example does not meet both conditions and is therefore approved.}
    \hfill
    \label{example section}
\end{figure*}

In implementing the \textsc{Rule System}, we drew inspiration from AutoMod, which allows community moderators to create automated scripts for detecting rule violations, such as profanity or external links. Recognizing that AutoMod’s complexity may deter end users from creating personal moderation classifiers, we used our pilot studies to develop a more user-friendly \textsc{Rule System} tailored to individual users' moderation needs (see Appendix A). The system filters comments that match any constructed rule, which represents a category of texts that the user wants to remove from their content feed. Figure \ref{rule block} presents an example of rules that users can create with our \textsc{Rule System}. Authoring rules is typically an iterative process in which users review examples, refine their rules, and examine the effect of their rules on those examples~\cite{dudley2018review, song2023modsandbox}. Therefore, we present examples from the training dataset on the side, as seen in Figure \ref{example section}, so that users can interactively review the effect of their rules and then adjust their rules accordingly.

\subsubsection{\textbf{\textsc{Prompt System}: Communicate with LLMs by Writing Prompts.}}

\begin{figure*}
    \centering
    \includegraphics[width=\textwidth]{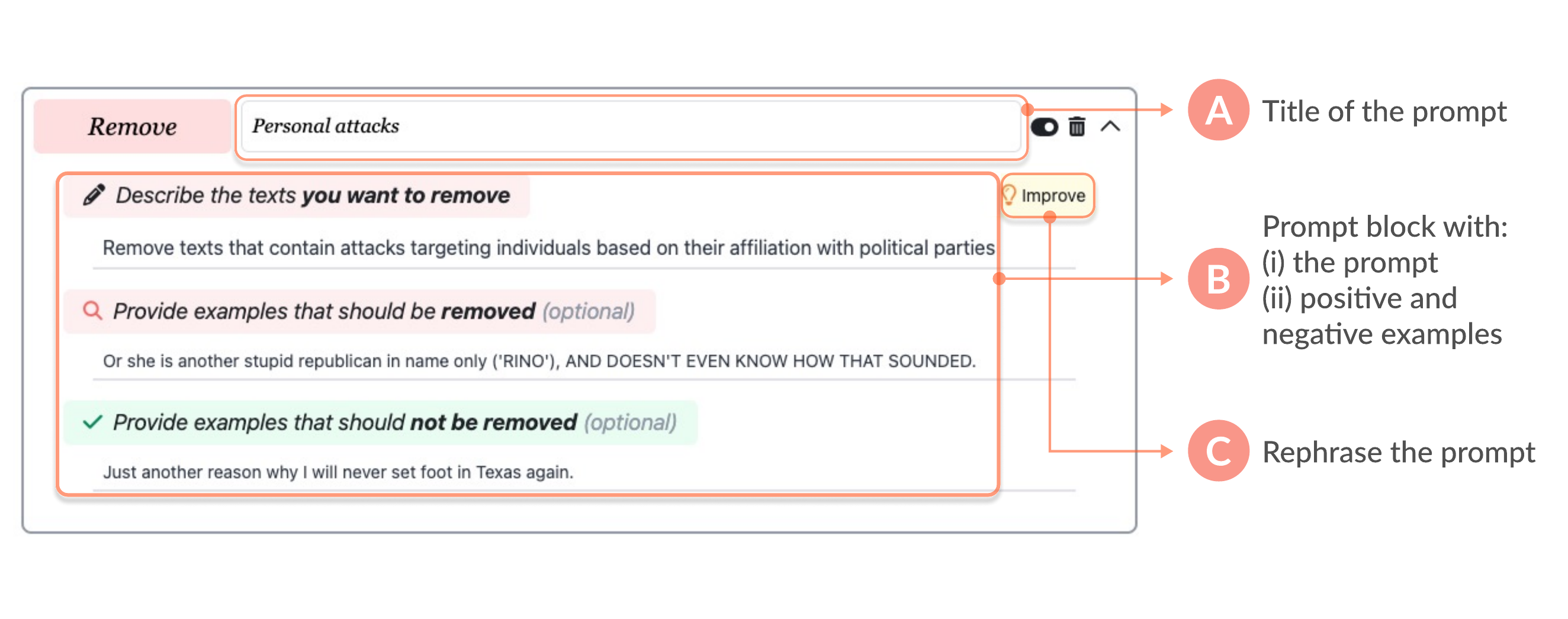}
    \caption{\textbf{An Example of a Prompt in the \textsc{Prompt System}}. Users describe the kinds of texts that they want to remove in natural language. Since few-shot learning is shown to significantly improve prompt performance~\cite{brown2020language}, users are encouraged to provide positive and negative examples. The LLM-backed ``improve'' feature rephrases users' descriptions to enhance LLM understanding.}
    \hfill
    \label{prompt block}
\end{figure*}

Like the \textsc{Rule System}, the \textsc{Prompt System} features a panel of instructions (prompts in this case) and a panel of examples. Figure \ref{prompt block} provides an example of a prompt as a category of unwanted comments. We refrain from asking LLMs to generate explanations for their predictions on individual examples because prior studies indicate that such explanations can be inaccurate, potentially confusing users rather than helping them develop clear mental models~\cite{turpin2024language, wang2023evaluating}. 

When developing the prediction algorithm, we needed to determine the optimal balance between the quality of LLM predictions and the time users must wait for predictions.
Through our pilot studies, we decided to prioritize low response times while maintaining sufficient prediction quality to reflect realistic deployment settings (see Appendix A).
Our prediction algorithm operates as follows. It combines each user-created prompt with a predefined system prompt and then requests predictions from LLMs in batches of 10 comments.
Since LLMs process each prompt independently, we can tell users which prompt leads to the removal of an example, thus offering more explainability than aggregating all prompts into a single query. We also introduced a caching mechanism\textemdash only re-querying the LLM for prompts that have been modified since their last evaluation\textemdash to minimize unnecessary requests and therefore reduce users' waiting times. Throughout our implementation, we used OpenAI's \texttt{gpt-4-1106-preview} model.

\subsection{Experiment Datasets}
For our experiment dataset, we crawled all comments of three videos about gun control policies from a YouTube channel called \textit{The Young Turks}: an online news show with 5.78 million subscribers that covers political topics from a left-leaning perspective. We refrained from using existing toxicity datasets~\cite{golbeck2017large, borkan2019nuanced, hada2021ruddit, chandrasekharan2018internet} because they were often sampled from various online platforms and communities to train more generic toxicity classifiers. Their examples relate to diverse contexts and are difficult for participants to understand. Hence, we concentrated on three videos regarding the same topic to simulate a real-world personal moderation setting more accurately.
We selected the channel \textit{The Young Turks} for various reasons. First, given our plan to recruit university students, we chose the familiar and opinion-provoking topic of political news over niche interest group content. Additionally, to elicit more opinionated reactions, we opted for a channel known for its more polarizing and occasionally controversial content. Lastly, we selected a channel without active comment moderation to ensure the presence of potentially toxic ones in our dataset.

We gathered over 5,000 comments from the three selected videos. To simplify the moderation task, we excluded comments that were responses to others. We also filtered out comments that were too brief or excessively lengthy, as they tended to be either irrelevant or cumbersome to read. Next, we used the Perspective API to assess the toxicity level of each comment. We labeled a comment as toxic if its toxicity score exceeded 0.7, following recommendations from prior research~\cite{hua2020towards, hua2020characterizing}. Although participants' moderation preferences can differ greatly, we aimed to balance the dataset so that nearly half of the comments would be toxic as determined by Perspective API. This process resulted in a balanced dataset of 800 comments. We randomly divided our dataset into a training dataset and a test dataset of 100 examples for each participant. The training dataset was used to help participants create their classifiers, whereas the test dataset was labeled by participants and used to evaluate their created classifiers.

\subsubsection{\textbf{Recruitment and Participants}}

We recruited 37 participants with no programming experience by advertising a call for participation on mailing lists of non-Computer Science departments at two major U.S.-based universities. We only selected participants who self-reported having little knowledge of programming and algorithms. There were 27 females, 9 males, and one participant who preferred not to disclose their gender. Most participants were pursuing their bachelor's degrees except four pursuing more advanced degrees.
Regarding political stance, 16 participants identified as liberal, 12 as moderate, and the remaining 9 were evenly distributed among ``very liberal,'' ``conservative,'' and ``prefer not to disclose'' groups. In terms of generative AI usage, 14 participants used it at least every few weeks, 14 participants every few months, and 9 participants rarely.

To gather qualitative data on participants' experiences with the systems, we conducted 13 individual user studies via Zoom. Additionally, we held 7 in-person workshops that did not include semi-structured interviews. Each workshop had between three to five participants, with all participants in a given workshop using the three systems in the same sequence. The individual user studies averaged 129 minutes in length, while the in-person workshops lasted about 100 minutes. Individuals were compensated with a \$40 gift card for their participation. In our analysis, we denote participants who attended individual sessions as P1--P13 and those who did not as W1--W24.

\subsection{Experimental Design}

\subsubsection{\textbf{Study Design and Procedure}}
We employed a within-subjects design with the three experiment systems described above. \leijie{We aimed to obtain at least 0.80 power to detect an effect size of 0.25 in precision at a standard 0.05 alpha error probability using F-tests. The required number of participants for each condition was 28~\cite{faul2007g}. Therefore, our experiment with 37 participants was sufficient for detecting the expected effect.}
The final experiment protocol was designed iteratively through five pilot experiments to ensure its effectiveness. 
This study was reviewed by our university IRB and deemed exempt. 

\textbf{Stage 1: Study Onboarding}. We started the experiment by briefing the participants and warning the possibility of encountering profanity and hate speech. We emphasized that participants could stop the experiment whenever they wanted, and we gained their explicit consent before proceeding. We then asked participants to imagine themselves as YouTube content creators whose videos on gun control policies had gone viral and were flooded with comments. They were then invited to set up automated classifiers to remove unwanted content based on their personal preferences. To gather a wide range of user preferences in our study, we assured participants that we would not judge their moderation preferences but instead focus on how well the three systems could align with their preferences.

\textbf{Stage 2: Ground Truth Labeling}. Subsequently, participants were asked to label 100 comments as ``Keep'' or ``Remove'' as the test dataset.
Prior research suggests that users might label data inconsistently, which harms the training of downstream ML algorithms~\cite{kulesza2014structured}.
Hence, we emphasized that participants should make labeling decisions consistently and that, if they change their criteria, they should revise their previous decisions. 
This test dataset would later be used to evaluate the performance of the classifiers that participants created.
We scheduled the ground truth labeling before the classifier creation so that participants could familiarize themselves with content moderation and this dataset of YouTube comments. While exposure to the test dataset beforehand might bias the classifier creation process, we argue that participants already had moderation preferences in mind and simply conveyed them when labeling their test datasets.

\textbf{Stage 3: Creating Personalized Classifiers}. We then asked participants to create classifiers using experiment systems in a randomized order. We used a counterbalanced design to counter any potential learning or fatigue effects. 
The underlying process for all three experiment conditions remained consistent. Each condition lasted 25 minutes in total. Participants first spent five minutes engaging with tutorial slides and the corresponding system. They were asked to try all of the system's functionalities and were encouraged to ask any questions. 

Then, in each condition, participants were given 15 minutes to create classifiers. 
In particular, the waiting time for backend computations did not count toward the 15-minute duration. This affected two systems: the \textsc{Prompt System} supported by a generative language model, and the \textsc{Label System}, which takes time to calculate the most uncertain examples for the next batch.
After 15 minutes, participants spent three minutes examining the overall performance and individual predictions of their created classifiers on their test datasets. Following this, participants reported their subjective experiences in a survey, which we will discuss in detail in \ref{evaluation}. Once participants completed all three conditions, we conducted a final survey to collect participants' preferred systems for content moderation and their rationales.

\textbf{Stage 4: Semi-structured Interviews (for Individual User Studies Only)}. 
At the end of the experiment, we asked participants a series of questions to understand the challenges that they encountered in creating and iterating on their classifiers. Examples of these questions include ``What do you like and dislike about writing prompts?'' and ``Can you easily understand why a prompt classifier makes its decisions?''
We also inquired about their preferred systems for various content curation scenarios.

\subsubsection{\textbf{Evaluation Measures}}
We consider the following three measures to evaluate the systems in our experiment.
\label{evaluation}
\textbf{Classification Performance}. To determine which system could learn user preferences with the highest performance, we evaluated the accuracy, precision, recall, and $F_1$ score of the final systems participants developed. 
We adopted several performance metrics because individuals prioritize different metrics for content moderation algorithms~\cite{wang2022ml, shen2022model}. Additionally, since participants' ground truth labels on the test dataset were not always balanced, accuracy might not reflect a system's true ability to distinguish between positive and negative examples.

\textbf{Creation Speed}. To assess which system could enable participants to develop a performant classifier most rapidly, we logged each participant's classifier every 30 seconds throughout the classifier creation period and calculated the performance of each snapshot. 
The performance of each classifier at early intervals serves as an indicator of its creation speed. In particular, we selected 5 minutes and 10 minutes as two representative intervals.

\textbf{Ease of Creation}. We documented 23 different types of user interactions, including asking for synonym suggestions, loading additional examples, and applying classifiers to examples.  These logged actions indicated the usefulness of the implemented features and how participants approached classifier creation with each system. 
We also gathered participants' subjective perceptions using a five-point Likert scale (from Strongly Disagree to Strongly Agree) for each condition:

\begin{itemize}
    \item \textit{Subjective Workload}. We adopted four applicable items from the NASA-TLX survey~\cite{hart1988development} regarding mental demand, temporal demand, effort, and feelings of stress. 

    \item \textit{System Usability}. We adopted all four items from the Usability Metric for User Experience (UMUX) survey~\cite{lewis2018measuring} to measure system usability. We chose the UMUX over the System Usability Scale (SUS) to reduce the overall number of questions that participants had to answer.

    \item \textit{Understanding}. To understand whether participants had a clear mental model of each system, we evaluated both participants' global understanding of a system and their local understanding of individual predictions~\cite{cheng2019explaining}. For global understandings, participants were asked to rate the statement  ``I felt that I had a good understanding of how my classifier works.'' For local understanding, we asked participants to explain the predictions of one false positive and one false negative randomly selected from the test dataset.
\end{itemize}

\begin{figure*}
    \centering
    \begin{subfigure}{\textwidth}
        \centering
         \caption{\textbf{Accuracy}}
        \includegraphics[width=\linewidth]{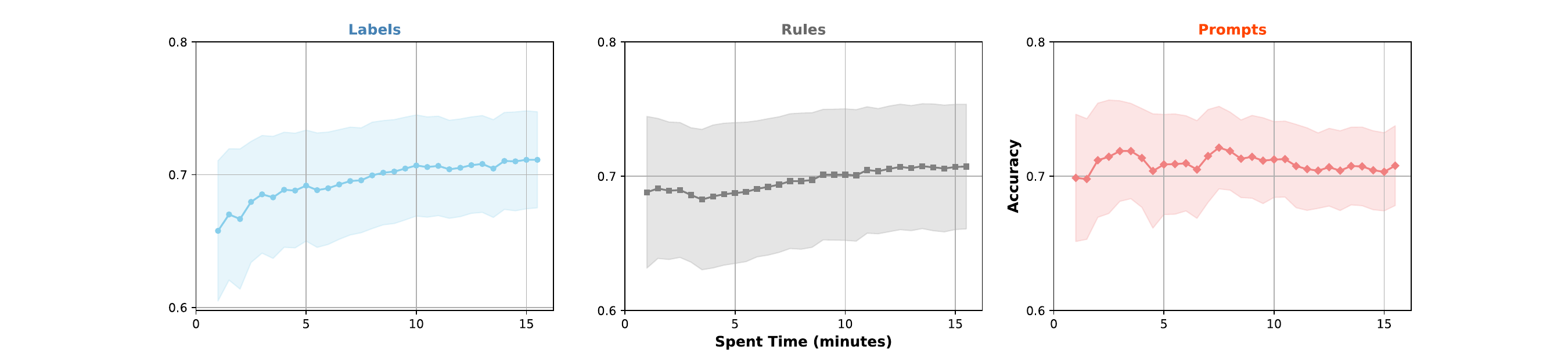}
        \label{fig:accuracy}
    \end{subfigure}
    \medskip
    \begin{subfigure}{\textwidth}
        \centering
        \caption{\textbf{Precision}}
        \includegraphics[width=\linewidth]{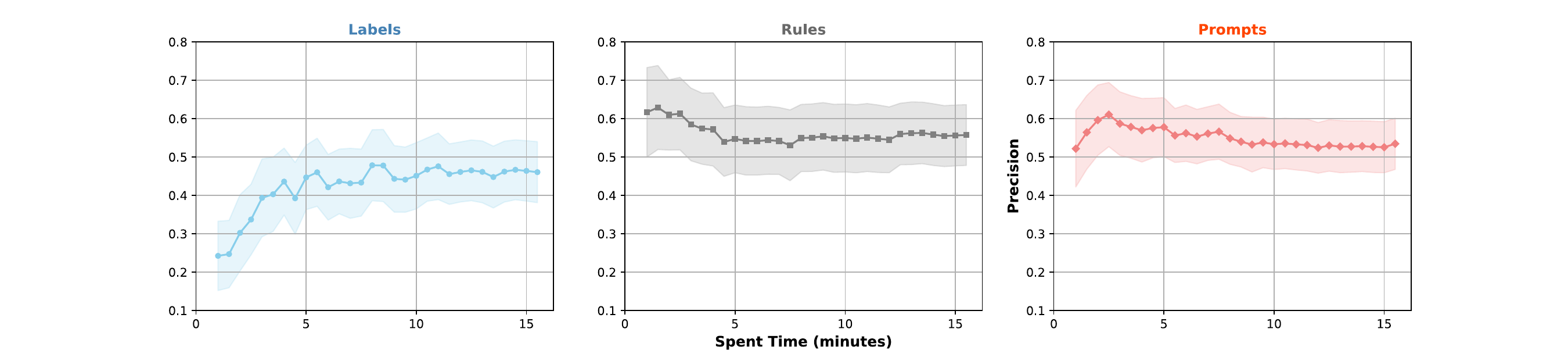}
        \label{fig:precision}
    \end{subfigure}
    \medskip
    \begin{subfigure}{\textwidth}
        \centering
        \caption{\textbf{Recall}}
        \includegraphics[width=\linewidth]{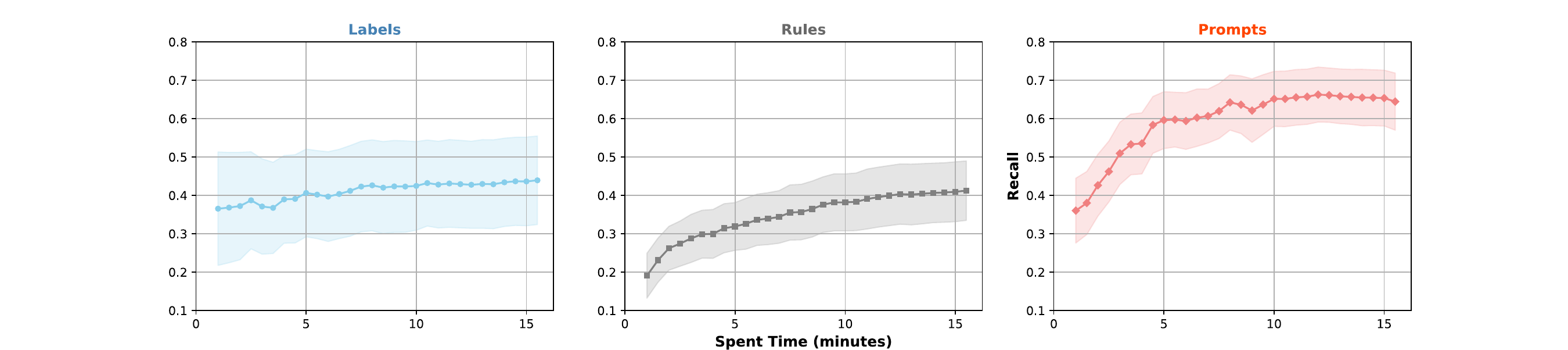}
        \label{fig:recall}
    \end{subfigure}
    \medskip
    \begin{subfigure}{\textwidth}
        \centering
        \caption{\textbf{F1 Score}}
        \includegraphics[width=\linewidth]{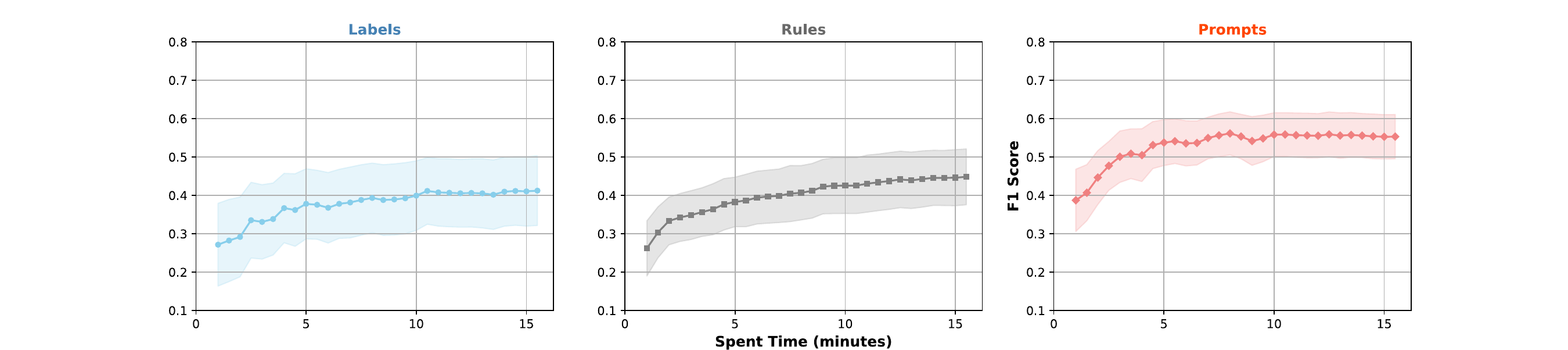}
        \label{fig:f1score}
    \end{subfigure}
    \caption{\textbf{Comparison of Performance Metrics Over Time Across Three Experimental Systems}. Each subfigure represents one of the following metrics: accuracy, precision, recall, and F1 score. Together, they illustrate how the average values of these metrics evolved from the first minute to the end of the study across all participants. We observed that the \textsc{Prompt System} has consistently higher recall and $F_1$ scores than the other two systems while maintaining comparable accuracy and precision. Moreover, while both the \textsc{Rule System} and the \textsc{Prompt System} exhibit increasing recall, these two systems experience plateaued precision and accuracy.}
    \label{performance}
\end{figure*}

\subsection{Data Analysis}
\subsubsection{Quantitative Modeling} Quantitative results were analyzed using a linear mixed-effects (LME) model, where experiment systems were treated as a fixed effect and participants as a random effect. The dependent variables in our model included classification performance, creation speed, and various subjective measures of user experience. We did not include the order in which participants used the three systems as another fixed effect, as no significant differences were observed for this variable. For each dependent variable, we calculated pairwise differences between the three experiment systems. \leijie{Finally, we provide basic information about the collected ground truth dataset. Each participant annotated 100 randomly sampled comments from the test dataset, with an average of 65 negative and 35 positive labels per participant.} We also conducted a sanity check to confirm that participants exhibited diverse moderation preferences. We found that half of the ground truth comments had, at most, a 75\% majority consensus.

\subsubsection{Qualitative Coding} Our qualitative data comprised semi-structured interview data and responses to open-ended questions in surveys. We employed a reflexive thematic analysis approach~\cite{Braun2019} to explore participants' experiences and challenges in creating personalized classifiers with each system. 
Reflexive thematic analysis has been widely used in HCI research to understand users’ experiences and views, as well as factors that influence particular phenomena or processes~\cite{Braun2019}. During data collection, the first author took detailed debriefing notes after each interview to document emerging themes. The authors then collectively reviewed the debrief notes and discussed themes in weekly group meetings. Recordings were automatically transcribed into text. The first author then open-coded the data on a line-by-line basis, and the remaining authors reviewed the transcripts and added codes. Over 300 codes were generated from the open-coding process. The authors clustered the open codes into high-level themes in a codebook and iteratively improved the codebook through discussion. Some examples of codes are \textit{Examples: Failed to understand preferences}, \textit{Rules: Transparency}, and \textit{Prompts: Idiosyncratic behaviors}. Finally, the authors applied the codes to the data to complete the thematic analysis.

\section{Results}
\subsection{What strategies enabled participants to create high-performing classifiers quickly?}

\subsubsection{The \textsc{Prompt System} resulted in the best final performance after 15 minutes, although both the \textsc{Prompt System} and the \textsc{Rule System} had the highest precision}.
Figure \ref{performance} compares the classification performances across the three systems over 15 minutes, with the endpoint of each line representing the final performance. Table \ref{effective models} presents the pairwise differences in the final performances among the three systems. We do not observe significant differences in accuracy between the systems. In terms of precision, \textsc{Rule System} showed significantly higher precision than \textsc{Label System} (Est.Diff = 0.097, $p$  < 0.05) but did not differ significantly from \textsc{Prompt System}. However, \textsc{Prompt System} showed significantly higher recall compared to both \textsc{Label System} (Est.Diff = 0.205, $p$  < 0.001) and the \textsc{Rule System} (Est.Diff = 0.232, $p$  < 0.001). Consequently, \textsc{Prompt System} also achieved a significantly higher $F_1$ score than the other two systems (Est.Diff compared to \textsc{Rule System} = 0.105, $p$  < 0.01; Est.Diff compared to the \textsc{Label System} = 0.141, $p$ < 0.001). 
We focus on $F_1$ as a performance metric of interest over accuracy due to the unbalanced nature of participants' labels.
\deleted{Finally, when examining how the final performance of each system varied across participants, we found that the \textsc{Prompt System} showed the least variance, regardless of the performance metric used. This suggests that the \textsc{Prompt System} is less susceptible to individual variances than the other systems.}

\subsubsection{The \textsc{Prompt System} reached 95\% of its peak performance within 5 minutes on average, with gains due to rapid initial improvements in recall}
Table \ref{effective models at 5} and Table \ref{effective models at 10} present the pairwise differences in classifier performance at 5 and 10 minutes respectively. Consistent with our earlier findings, the \textsc{Prompt System} demonstrated significantly higher recall and $F_1$ scores than the other two systems.
\deleted{Notably, its recall at 5 minutes had already surpassed the final recall of the other two systems.}
Meanwhile, the \textsc{Prompt System}'s recall exhibited a distinct pattern: a rapid increase in the first five minutes, followed by a steady climb in the subsequent five minutes, and a plateau in the remaining time. In contrast, the \textsc{Rule System} showed a consistent increase in recall throughout the entire 15-minute period, whereas the \textsc{Label System} showed a less noticeably upward trend in recall over time.
On average, participants reached 95\% of their peak recall by 7 minutes, and that of their peak $F_1$  score by 5 minutes. We also observed that participants made fewer changes to their prompts as time went on, while they continued actively labeling examples or curating keywords for their rules in the other two systems. These findings suggest that writing prompts can facilitate rapid initialization but may be less effective in supporting iterative improvements.

\subsubsection{Users struggled to improve the precision of all three systems over time.}

The precision of the \textsc{Prompt System} and the \textsc{Rule System} remained comparable throughout the experiment. 
\deleted{While the \textsc{Prompt System} showed significantly higher precision than the \textsc{Label System} at 5 minutes (Est.Diff = 0.131, $p$ < 0.05), the difference became non-significant at 10 minutes (Est.Diff = 0.082, $p$ = 0.090). In contrast, while initially comparable at 5 minutes (Est.Diff = 0.100, $p$ = 0.06), the \textsc{Rule System} had a significantly higher precision than the \textsc{Label System} at 10 minutes (Est.Diff = 0.099, $p$ < 0.05) and at the end (Est.Diff = 0.097, $p$ < 0.05).}
While both the \textsc{Prompt System} and the \textsc{Rule System} showed notable improvements in recall by the end of the experiment, their precision either plateaued or even decreased over time. 
For instance, the \textsc{Rule System} experienced a gradual decrease in precision during the first 5 minutes, followed by a steady state for the remaining 10 minutes. The \textsc{Prompt System}, despite a slight increase in precision in the initial 2 minutes, subsequently experienced a minor decline before reaching a plateau. 
These observations align with participant behaviors: participants tended to add more categories of unwanted content rather than increase the specificity of existing rules or prompts during the task because the latter was often considered more cognitively demanding.
Although the \textsc{Label System} demonstrated a rapid increase in precision during the first five minutes, it began with such low precision that its final precision was still below that of the other two systems.

\subsubsection{Summary}
In summary, the \textsc{Prompt System} enabled end users to create custom moderation classifiers with superior performance more rapidly compared to the other two systems. 
In terms of the precision-recall trade-off, its higher performance was primarily driven by its highest recall, as its precision remained comparable to the \textsc{Rule System}. 
Temporally, we observed that the \textsc{Prompt System} facilitated rapid classifier initialization during the early stage but faced challenges in supporting further iteration, as evidenced by its plateaued precision and recall afterward.
In contrast, both the \textsc{Rule System} and the \textsc{Label System} showed steadily increasing trends in their recall and $F_1$ scores throughout the experiment, despite being constrained by their significantly lower initial values.

\begin{table*}[ht]
\centering
\captionsetup[table]{skip=10pt}
\begin{tabular}{l|cc|cc|cc|cc}
\toprule
& \multicolumn{2}{c|}{\textbf{Accuracy}} & \multicolumn{2}{c|}{\textbf{Precision}} & \multicolumn{2}{c|}{\textbf{Recall}} & \multicolumn{2}{c}{\textbf{$F_1$  Score}} \\

Pairwise Comparison & Estimate & Std. Err. & Estimate & Std. Err. & Estimate & Std. Err. & Estimate & Std. Err. \\
\hline
Rules - Labels & -0.004 & 0.019 & 0.097* & 0.040 & -0.027 & 0.053 & 0.036 & 0.038 \\
Prompts - Labels & -0.003 & 0.019 & 0.074 & 0.040 & 0.205*** & 0.053 & 0.141*** & 0.038 \\
Prompts - Rules & 0.001 & 0.019 & -0.023 & 0.040 & 0.232*** & 0.053 & 0.105** & 0.038 \\
\bottomrule
\end{tabular}
\hfill\\
\hfill\\
\caption{\textbf{Post-hoc pairwise t-test results for the final performance}: This table displays pairwise differences in mean performance metrics of the final classifier that participants developed using the three systems. $p$ < 0.001***, $p$ < 0.01**, $p$ < 0.05*}
\label{effective models}
\end{table*}

\begin{table*}
\centering
\captionsetup[table]{skip=10pt}
\begin{tabular}{l|cc|cc|cc|cc}
\toprule
& \multicolumn{2}{c|}{\textbf{Accuracy}} & \multicolumn{2}{c|}{\textbf{Precision}} & \multicolumn{2}{c|}{\textbf{Recall}} & \multicolumn{2}{c}{\textbf{$F_1$  Score}} \\

Pairwise Comparison & Estimate & Std. Err. & Estimate & Std. Err. & Estimate & Std. Err. & Estimate & Std. Err. \\
\hline
Rules - Labels & -0.004 & 0.021 & 0.100 & 0.053 & -0.087 & 0.059 & 0.005 & 0.043 \\
Prompts - Labels & 0.017 & 0.021 & 0.131* & 0.053 & 0.190** & 0.059 & 0.159*** & 0.043 \\
Prompts - Rules & 0.021 & 0.021 & 0.031 & 0.053 & 0.277*** & 0.059 & 0.155*** & 0.043 \\
\bottomrule
\end{tabular}
\hfill\\
\hfill\\
\caption{\textbf{Post-hoc pairwise t-test results for the classifier performance at 5 minutes}: This table displays pairwise differences in mean performance metrics of classifiers developed by participants during the first 5 minutes using the three systems. $p$ < 0.001***, $p$ < 0.01**, $p$ < 0.05*}
\label{effective models at 5}
\end{table*}

\begin{table*}
\centering
\captionsetup[table]{skip=10pt}
\begin{tabular}{l|cc|cc|cc|cc}
\toprule
& \multicolumn{2}{c|}{\textbf{Accuracy}} & \multicolumn{2}{c|}{\textbf{Precision}} & \multicolumn{2}{c|}{\textbf{Recall}} & \multicolumn{2}{c}{\textbf{$F_1$  Score}} \\

Pairwise Comparison & Estimate & Std. Err. & Estimate & Std. Err. & Estimate & Std. Err. & Estimate & Std. Err. \\
\hline
Rules - Labels & -0.006 & 0.020 & 0.099* & 0.048 & -0.043 & 0.051 & 0.026 & 0.039 \\
Prompts - Labels & 0.006 & 0.020 & 0.082 & 0.048 & 0.227*** & 0.051 & 0.159*** & 0.039 \\
Prompts - Rules & 0.011 & 0.020 & -0.016 & 0.048 & 0.270*** & 0.051 & 0.133*** & 0.039 \\
\bottomrule
\end{tabular}
\hfill\\
\hfill\\
\caption{\textbf{Post-hoc pairwise t-test results for the classifier performance at 10 minutes}: This table displays pairwise differences in mean performance metrics of classifiers developed by participants during the first 10 minutes using the three systems. $p$ < 0.001***, $p$ < 0.01**, $p$ < 0.05*}
\label{effective models at 10}
\end{table*}

\subsection{What strategies did participants find easiest for communicating their preferences?}

\begin{figure*}
    \centering
    \includegraphics[width=\textwidth]{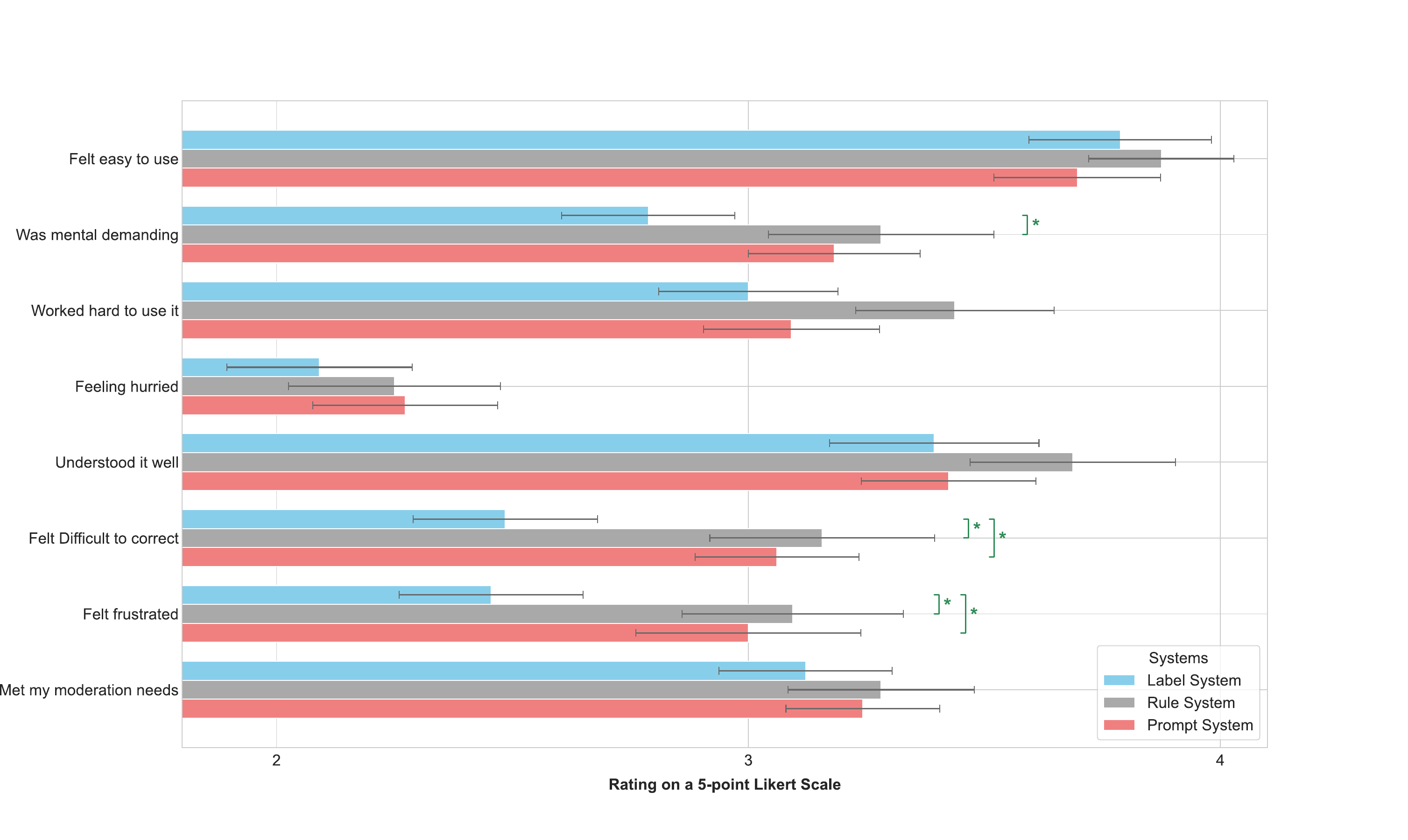}
    \caption{\textbf{Comparison of Subjective Measures Across the Three Systems}. Participants rated their experiences in terms of task load and usability. This figure displays the comparison of their ratings across the three systems. Error bars indicate standard deviation, and p-values are provided to show statistical significance. We found that participants considered the \textsc{Rule System} and the \textsc{Prompt System} to be more mentally demanding and frustrating than the \textsc{Label System}. We did not find significant evidence that participants deemed the \textsc{Prompt System} easier to use or better at meeting their moderation needs compared to the other two systems.}
    \hfill
    \label{usability}
\end{figure*}

\subsubsection{When participants had ill-defined but intuitive preferences, they found the \textsc{Label System} to be easiest}
Unlike professional moderators, social media users often lack the time or opportunity to articulate their moderation preferences clearly~\cite{jhaver2023personalizing}. Instead, their moderation decisions are often a result of their intuitive feelings about the content they encounter. 
\deleted{P4 pointed out that ``\textit{sometimes it might be the entire sentence that I don't want to see, or maybe it's just about a word...It could also be the feeling of the comment. You don't feel right about it.}''}
Some participants also lacked ``\textit{a holistic view about what were the alarming comments}'' (P2), since new content continuously flooded their social media feeds. 
Labeling examples thus provided them with a natural way to express their intuitive preferences to the algorithm, whereas writing prompts or authoring rules forced them to distill their intuitive preferences into high-level patterns.
Participants tended to trust algorithms to infer their preferences from their labels more than they trusted themselves to accurately convey their preferences, especially in time- and attention-constrained settings of personal moderation.
As P2 explained, ``\textit{I would actually trust the algorithm finding trends more [than] me looking at everything and trying to build my own...especially with my lack of experience and lack of time.}'' 
\deleted{This trust made participants more willing to offload the task of summarizing high-level criteria to the algorithm, despite this requiring them to label many examples.}

In contrast, translating their intuitive preferences into rules or prompts required participants to iteratively refine their rules or prompts to better convey their intuitions, a process that participants largely considered to be mentally demanding. P10 described their process of writing prompts in detail: ``\textit{In the beginning, I have an idea of what I do not want to see and then I would submit that. But [AI] wouldn't know what I was talking about, so I had to change my wording to match what AI would understand.}'' 
For some participants, this process nudged them to actively reflect on their intuitions about what kinds of content they wanted to moderate, but for others, it was simply too mentally taxing. P2 expressed the latter sentiment: ``\textit{It was almost more tiring to write prompts [than label examples], as you need to think about each one but then apply it, and then see one little thing you could have fixed, and then reapply it.}'' This is part of the reason why participants rated the \textsc{Label System} as significantly less frustrating and demanding than the other two systems in Figure \ref{usability}.


\subsubsection{When participants had well-defined and general preferences, they found the \textsc{Prompt System} to be easiest}
Some participants could clearly define their general preferences regarding themes like violence or hate speech before even engaging with a classifier.
Unlike authoring rules, they could directly translate those preferences into prompts without the need to curate a comprehensive list of keywords. 
P6 explained this advantage of writing prompts: ``\textit{For rules, if there are 1,000 words I don't want to see, I have to list all 1,000 words, [whereas for ChatGPT], I only need to think of 10 words among those 1,000 words, and then ChatGPT itself understands that this guy does not want this kind of words.}'' 
This advantage is especially pronounced because some preferences are hard to capture with keywords. P10 highlighted their struggle to author rules for such preferences: ``\textit{It is easier to think about specific words for removing profane content, but then for violence, I didn't know what words specifically to say [for such content].}''
Even if participants could identify keywords to indicate their preferences, they sometimes hesitated to add them because these keywords could also be used in neutral or benign contexts and thus adding them could result in many false positives.
Additionally, writing prompts simplified the articulation of complex preferences, which would otherwise require complex rule structures.
For instance, P11 tried to create a rule to ``\textit{remove personal attacks against an organization.}'' Even though they managed to compile two lists of phrases---one for ``personal attacks'' and another for ``organizations''---they found that the conjunction of these two concepts in a rule structure did not accurately capture the intended relationship between them.

While rules provided participants with a structured, albeit somewhat cumbersome, way to map their pre-established preferences, labeling examples was considered the least straightforward method for expressing such preferences. 
P13 communicated their frustration with the inability to explicitly state their preferences through labeling: ``\textit{There are too many decisions to make. Yeah, they're small decisions of yes or no. But usually, I don't have to keep creating 50 small decisions to indicate my personal preferences. Why am I going through hundreds of these when I just want to focus on saying my preferences?}''
Participants' senses of frustration further increased when they could not find sufficient relevant examples to label as a way of indicating their preferences, as described by P10: ``\textit{I feel like I have this criterion [remove violence calls] in my head. But there is nowhere I can label these examples.}''
This issue is especially severe when users try to develop a classifier for a new community with a limited number of examples available.
Finally, participants were concerned about whether algorithms would accurately learn their high-level preferences from labeled examples. P13 worried, ``\textit{What if all the comments happened to have [the word] Texas, so the AI just picks up those all related to Texas?}''

Despite the prompts' advantages over the other approaches, participants still found it challenging to describe concepts outside of LLMs' existing knowledge. For example, LLMs lacked awareness of the context, such as the video or post to which comments were attached. P7 wanted to ``\textit{remove irrelevant comments}'' and wished for ``\textit{ChatGPT to be familiar with the content of the original post.}'' Additionally, LLMs often struggled with more complex concepts such as misinformation or conspiracy theories. W14 experienced this limitation: ``\textit{I wanted to remove this comment because it contains obvious disinformation about the city of Portland Oregon, but my filter did not understand my definition of obvious disinformation.}''


\subsubsection{When participants had well-defined preferences regarding specific topics or events, they found the \textsc{Rule System} to be easiest}

While rules as a keyword-based approach may overlook the broader context of comments, they can effectively adhere to preferences that can be captured by a list of keywords. Although people could write prompts that resemble rules in the \textsc{Prompt System}, these prompts could still remove comments that included similar concepts. P8 described the difference between authoring rules and writing prompts along this dimension: ``\textit{[By authoring rules], you might get to really specific thing, and lost the big picture. But then [by writing prompts], you have just these larger generalizations. You know you can't get down to the nitty-gritty and identify truly exactly what phrases you want to remove.}'' In addition, rules are particularly useful for hiding content about specific events. As P1 explained, ``\textit{I would only use the rule system if there was a particular topic that was very distressing for me, like a bombing somewhere. If I were personally affected by that and didn't want to hear anything about that, I would just create a short list of relevant words.}'' 

\subsection{What strategies were easiest for iterating on classifiers after initialization?}

\subsubsection{Transparency in the \textsc{Rule System} and the \textsc{Prompt System} enabled participants to pinpoint problems but not always be able to fix them, whereas the \textsc{Label System} offered less transparency and little opportunities for targeted iteration.}

Participants had a clearer understanding of why the \textsc{Rule System} made incorrect predictions than the other two systems. Such transparency helped participants improve their classifiers in some cases, like when they forgot to include spelling variants or similar phrases to those that they explicitly included in their rules. But in other cases, transparency did not lead to iterative improvements because rules could not accommodate contextual variances, as documented in prior research~\cite{song2023modsandbox, reddit2023automod}. Our survey and interviews suggest that participants had a slightly better understanding of why their classifiers made mistakes for the \textsc{Prompt System} than for the \textsc{Label System} (Figure \ref{usability}). Participants found that they could at least review and reflect on problematic prompts in the \textsc{Prompt System}, whereas they could not pinpoint specific labels that led to classification mistakes in the \textsc{Label System}. 
\deleted{ P2 noted this advantage: ``\textit{The prompt system was more transparent in the sense that it highlights specific guidelines you have set up. You could then change them and see what they do.}'' 
For the \textsc{Label System}, participants had only a vague understanding of why their classifiers made incorrect predictions, reflecting previous research on how end users develop folk theories to interpret content curation algorithms~\cite{eslami2016first, mayworm2024content, devito2017algorithms}.} 
Moreover, the \textsc{Label System} offered few opportunities for targeted iteration, which were particularly useful as participants found it easier to point out what the classifier should learn from its mistake. This highlights the need for more flexible approaches at different stages of classifier creation, as recent studies advocated~\cite{piorkowski2023aimee, daly2021user}. 

\subsubsection{Users of the \textsc{Prompt System} struggled to refine their initial prompts due to human-LLM misalignment and LLMs' unpredictable behaviors.}

While participants could quickly build classifiers with decent performance using the \textsc{Prompt System}, they found it challenging to incorporate more nuances when iterating on their prompts. Participants rated it as comparably difficult to correct with the \textsc{Rule System} (Figure \ref{usability}).
Many participants described a disconnect in how humans and LLMs perceive the same prompts. They observed that ``\textit{descriptions could be very subjective}'' (P4) and asked, ``\textit{What has [the LLM] previously been trained on? For words they labeled as extreme, I might not think as extreme}'' (P13). 
Consequently, many noticed that LLMs seemed to interpret their preferences broadly, causing LLMs to have trouble distinguishing different degrees of a concept.
For instance, P13 noted that ``\textit{[LLMs] have the problem of differentiating between just minor attacks on someone's character [such as calling someone a fool] and ones that are strong and can be perceived as harm.}''
Similarly, P8 struggled to teach LLMs to ``\textit{differentiate between threats and just using the word `shoot' or `kill.'}'' Articulating the fine-grained differences between concepts proved too difficult for many participants.

In addition, participants sometimes faced confusion in response to the peculiar behaviors of LLMs, leaving them unsure of how to further improve their prompts. 
For instance, P3 was perplexed by their prompt classifier's removal of the comment ``\textit{Jessica is the best},'' explaining, ``\textit{I'm honestly not sure why the filter removed this content. I wrote something about removing comments with harmful descriptions, so maybe it thought that this description was harmful in some way.}''
Sometimes, participants were surprised that the \textsc{Prompt System} did not remove comments that they included as few-shot examples. P1 frustratedly said, ``\textit{With ChatGPT, I would directly copy and paste a comment that I didn't want, and it still wouldn't remove it.}'' 
Even though the performances of participants' prompt classifiers were comparable to the other two systems' performances, such incomprehensible behaviors undermined their confidence in using LLMs for personal moderation. 

\subsubsection{Even when using the \textsc{Prompt System}, participants iterated using \textsc{Rule System}- and \textsc{Label System}-like strategies}

Surprisingly, to further refine their prompts to align with their nuanced preferences, participants often learned from strategies of authoring rules or labeling examples. Several participants opted to avoid using high-level descriptions in their prompts and instead wrote prompts that resembled rules, such as ``\textit{Remove texts that refer to people as stupid, dumb, idiots [a list of words].}'' In this way, they expected that LLMs could still catch various spelling variants of these words but would not generalize more broadly beyond the word list.
Alternatively, when unsure how to describe nuances in prompts, many participants resorted to adding representative positive or negative examples directly into their prompts as few-shot examples. In this process, they were essentially labeling examples rather than writing prompts. Four participants even included more than five examples in their prompts. 
W6 remarked: ``\textit{I liked that I could copy and paste certain examples in ChatGPT. It made it easier to capture ideas that specific words do not allude to but are alluded to by unique comments.}'' 
However, some participants also recognized the limitations of this approach. As P13 noted, ``\textit{I think [adding examples to prompts] is good up to a certain extent. 
If you know there are comments you don't want, it's great and...also gives you more nuances. 
But sometimes when a lot of these comments are extremely similar, [the LLM] just gets confused.}''


\begin{figure*}
    \centering
    \includegraphics[width=0.7\textwidth]{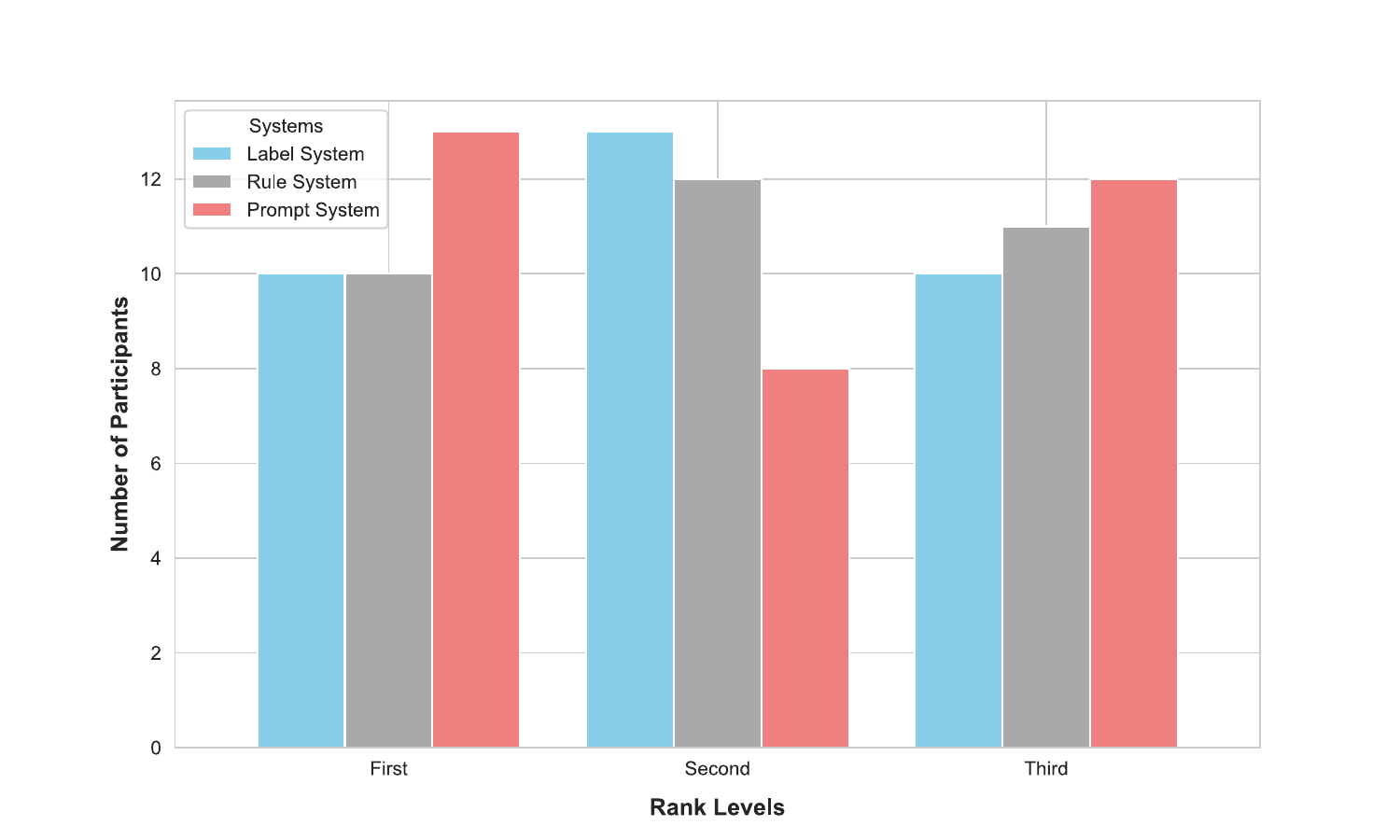}
    \caption{\textbf{Participant Preferences for Personal Moderation Systems}. This figure shows how participants ranked the three systems in order of preference for use in real-life personal moderation. Results from the Wilcoxon Signed-Rank test indicated no significant differences between any pair of systems.}
    \hfill
    \label{preferences}
\end{figure*}

\subsection{Which strategies did participants prefer overall?}

\subsubsection{There was no clear preference for a particular system across all participants; instead, different systems were favored based on varying use cases and individual needs.}
We asked participants to rank the three systems regarding their preferences for using them in real-life content moderation and to explain their rationale for these rankings. 
Figure \ref{preferences} illustrates the distribution of participants' rankings. 
Surprisingly, despite growing enthusiasm for using LLMs to moderate content, a comparable number of participants still preferred labeling examples or authoring rules to create custom classifiers for content moderation. 
The Wilcoxon Signed-Rank test showed no significant differences between any of the systems. 

By examining data from our semi-structured interviews and open-ended responses from the final survey, we identified several dimensions that contribute to this variance in preferences. 
They highlight the principal differences across the three systems, the diversity of individual preferences, and the variety of content curation scenarios for end users.

\begin{itemize}[leftmargin=*]
    \item 

\textbf{Whether users want explicit articulation of their preferences.} Individual preferences for content curation are nuanced and subject to change. 
As a result, if users had explicitly stated their preferences in rules or prompts, they would have to continually reflect on their preferences and update their classifiers accordingly. In comparison, the \textsc{Label System}'s algorithms had the potential to learn these subtle changes in preferences from a continuous flux of labels. As P12 explained, ``\textit{Preferences change over time, and we may not explicitly realize that. So you may accidentally remove some things if your preferences change but you have already explicitly set them in stone. }'' However, some participants appreciated the chance to actively reflect on and articulate their preferences through high-level rules or prompts.


\item
\textbf{Whether users can tolerate reviewing many toxic examples.} In our study, participants reviewed fewer comments to develop an effective classifier using the \textsc{Prompt System} compared to the other two systems. On average, participants reviewed 133 and 127 examples for the \textsc{Label System} and the \textsc{Rule System} respectively, whereas only 53 examples were needed for the \textsc{Prompt System}. The need to review numerous toxic examples could be distressing, particularly for those creating classifiers to block such content. Whether people are willing to undertake such an emotional toll varies across moderation scenarios. As P8 stated, ``\textit{If I were moderating for my own sake and not for a community, I would probably just focus more on the ease of creating a filter so that I don't have to subject myself to these potentially offensive comments.}'' 

\item
\textbf{Whether users prioritize precision over recall.} Participants demonstrated varied approaches to balancing precision and recall. Some favored precision because they were more concerned about the risk of removing benign comments than approving unwanted content. P12 described their rationale as follows: ``\textit{
I don't want to fall into any echo chamber. Even if there are really crazy people saying really crazy things, they make me frustrated, but I still should know about it.}'' 
Other participants favored high recall, even if it meant mistakenly removing benign content. 
P13 expressed this preference: ``\textit{
Personally, too specific is worse than the too broad. 
If the filter keeps certain content I really don't want to see, that's far worse than if I didn't get an extra post. Because of how much content there already is, I'll just get others on my feed anyway.}'' 
For these participants, the high recall of the \textsc{Prompt System} is even advantageous during the creation stage. Since LLMs could collect the majority of potentially unwanted examples given initial criteria, participants could focus on reviewing these examples to increase the precision of their prompts.

\item
\textbf{To what extent users value transparency and controllability.} An important distinction between the \textsc{Rule System} and the other two systems lies in transparency. 
As P10 described, ``\textit{I like rules more because I know what's going into them. But with the prompts, you don't know what's going on. While the prompting is better at catching things, I just like the transparency of rules more.}''
The \textsc{Rule System} helps users not only understand why exactly a comment is removed; it also enables moderators and users to have clear expectations about which comments will be removed. For users who moderate for a community, this sense of controllability enables them to be more accountable for their moderation decisions.
\end{itemize}


\section{Discussion}

In this work, we compared three prominent strategies for creating custom classifiers, focusing on their support for rapid initialization and easy iteration. Our experiments revealed that, despite its high performance, writing prompts also suffered from challenges in further iteration and lack of transparency. In this section, we discuss how incorporating labeling examples and authoring rules could help mitigate these problems, and how the unique user needs of diverse content curation applications demand different hybrid systems.

\subsection{Hybrid Approaches to Personalized Classifier Creation}
Existing tools often fail to accommodate the diversity and fluidity of user needs in content curation. Our experiment indicates the potential to incorporate labeling examples and authoring rules into future content curation systems, thereby providing end users with more flexible strategies to create and iterate on their classifiers. 

Labeling examples can facilitate easier \textit{prompt bootstrapping}. 
Our findings indicate that when users have ill-defined but intuitive preferences, they prefer labeling examples over writing prompts to convey those preferences. 
However, with traditional ML-based algorithms, users often complain that they need to label numerous examples to express their preferences and that there is a lack of transparency regarding what the algorithm learned from their labels.
We envision that users can provide a few representative examples for LLMs, which help infer and suggest potential preferences.
Labeling examples could also help users \textit{easily iterate} on their prompts. In our experiments, when users tried to refine their prompts, they often focused on a few misclassified examples and adjusted their prompts until LLMs made correct predictions.
Given the mental load of such interactive iteration for many participants, we propose developing an automated prompt chain where LLMs automatically refine prompts based on a few corner-case examples users curate and label.

Authoring rules offers users a stronger sense of \textit{transparency and controllability} than writing prompts. Our experiments indicate that rules could easily capture preferences involving specific topics or events.
Therefore, instead of completely replacing rules with prompts, content curation tools should allow users to choose between rules and prompts based on their preferences.
Future research should also investigate ways to increase the transparency of LLM predictions. 
In our implementation of the \textsc{Prompt System}, we asked LLMs to process each prompt independently rather than aggregating all prompts into a single query. 
This approach is akin to connecting prompts in a rule structure, thus offering more transparency. 
Similarly, researchers could introduce more transparency into LLM predictions by decomposing a complex prompt into a series of conditions and querying LLMs separately. 
This method allows users to understand which specific conditions might be causing mistakes~\cite{zhou2022least, dua2022successive}. 

\subsection{Supporting Different Content Curation Applications}
As discussed, content curation scenarios can range from curating personal feeds to managing community content. Individuals could also have intuitive preferences or gradually develop more well-defined preferences from initial intuitions. Additionally, individuals vary greatly in many dimensions, such as tolerance of undesirable content and the precision-recall trade-off. Future research should therefore only apply our findings after closely examining the unique needs of their contexts.
For example, individuals looking to remove undesirable content from their feeds may prefer prompt-based systems due to their ease of use. In contrast, prompt-based systems might be less desirable for community moderators without comparable transparency to existing rule systems, as moderators are more accountable for their moderation decisions~\cite{song2023modsandbox}. 
Additionally, despite their algorithmic similarities, curating desired content requires different systems than removing undesirable content, since users tend to prioritize recall over precision during active exploration and have more fluid and intuitive preferences than in content moderation~\cite{feng2024mapping}.

In this work, we investigate supporting users to create personal classifiers \textit{from scratch} but social media users might want to \textit{modify existing classifiers}. Some platforms have already offered similar but simplified functionalities, allowing users to adjust the sensitivity for predefined concepts like racism and misogyny~\cite{Instagram2024, bhargava2019gobo, Bodyguard2024}. Compared to a platform-centric definition of these concepts, end users might feel more comfortable customizing classifiers created by friends or family members who share similar interests and moderation preferences~\cite{mahar2018squadbox}. 
Our findings also advocate for more flexible customization options beyond simply adjusting thresholds~\cite{jhaver2023personalizing}. 
For example, we envision users being able to merge labeled data, fork and edit rules, or share prompts. \leijie{However, our findings might be less applicable here because users focus less on rapidly initializing a classifier and more on understanding the decision-making boundaries of existing classifiers. In such cases, sharing prompts and rules might be preferred over merging labeled examples, as their high-level nature facilitates quicker understanding.}

While we focus on content-based curation in this work, end users also frequently curate content based on metadata. 
For instance, they might want to see all posts from certain accounts regardless of content or may prefer the latest content when curating time-sensitive information. Here, the challenge lies in helping users curate and summarize information for decision-making, such as evaluating whether a user is worth following or identifying when content becomes interesting. Future research should explore how to support end users in communicating preferences that involve conflicting non-textual and textual information (e.g., what if someone I follow posts unwanted content?).
Additionally, while our findings on binary classifiers could be generalized to a categorical classifier, they cannot be easily extended to regression algorithms. 
Such algorithms are still valuable for users who want to sort content in their feeds, as in the context of content recommendation~\cite{adomavicius2005toward}. 
Common content recommendation systems often learn from example-level feedback but suffer from similar issues identified in our experiments, such as few opportunities for explicit feedback and lack of transparency~\cite{feng2024mapping}. Future work should investigate the potential of using rules or prompts in this space.

Finally, while there is growing interest in more cost-effective LLMs~\cite{chen2023frugalgpt, yue2023large, vsakota2024fly}, the cost of deploying an LLM-based content curation system remains a concern given the vast amount of online content produced daily. 
For instance, on a platform level, there are 500 million tweets sent every day~\cite{TwitterDay}, and on a user level, a popular YouTube video can gather thousands of comments.
Although techniques like multi-step reasoning~\cite{dua2022successive} or self-consistency~\cite{wang2022self} can enhance the performance of LLMs on complex user preferences, they often require more computational resources. Future research should investigate leveraging techniques such as LLM cascading~\cite{chen2023frugalgpt} or model distillation~\cite{xu2024survey} to maintain performance while reducing costs. For example, cheaper models could be used for less complex or less important user preferences, whereas LLMs could be distilled into lightweight classifiers for heavy users.

\section{Limitations}
While our experiment protocol was developed iteratively through pilot studies, it still has limitations. First, we determined our three experiment systems by mapping each strategy to their most common backend algorithms (supervised learning algorithms for labels, transparent algorithms for rules, and LLMs for prompts). As a result, we did not have the chance to test other less common combinations in our experiments, such as supervised learning algorithms for rules (i.e., a Snorkel-like system~\cite{ratner2017snorkel}).
Second, even though we explicitly encouraged participants to label the test dataset consistently, exposure to more examples from the training dataset during classifier creation might alter their criteria. To prevent the experiments from being excessively long, we limited participants to labeling only 100 comments that comprised our test dataset, which might affect our evaluation's accuracy. 

Besides, our participant pool of undergraduates did not represent the diverse population of internet users, and our study involved a relatively small number of participants. Future work should conduct larger-scale experiments to validate our findings.  
Fourth, our study was conducted solely in English and focused on political content, limiting its applicability to other languages and types of content. Since LLMs might have different performances for other languages or content types~\cite{shen2024language}, future research should validate and extend our findings.
Finally, while personal content moderation as a case study illustrates the primary user needs for content curation tools, we acknowledge the diversity of content curation scenarios. Future work should examine whether there are distinct user needs in other scenarios before applying our findings broadly.

\section{Conclusion}
\leijie{Existing interactive machine learning systems often make assumptions about end users that do not translate well into the context of content curation.
These systems often assume that users are highly motivated to train an algorithm in a single, lengthy session. 
In contrast, social media users prefer to build their classifiers casually, engaging in shorter, iterative sessions as part of their daily routines.
This raises a key question: how can we better support end-users in easily bootstrapping and iterating on classifiers for content curation?
To address this, we compared three prominent strategies for creating custom content classifiers: \textit{labeling examples for supervised learning}, \textit{writing and carrying out rules}, and \textit{prompting a large language model (LLM)}. From a within-subjects experiment with 37 non-programmer participants, we found that participants preferred different strategies based on their moderation preferences, despite relatively higher performances of prompting LLMs.}
While prompts could effectively communicate users' well-defined and general preferences, participants preferred labeling examples to convey their ill-defined but intuitive preferences and authoring rules for their preferences about specific topics or events.
Moreover, participants found it challenging to refine their prompts to communicate their nuanced preferences iteratively. Consequently, they often directly added misclassified examples as few-shot examples or attempted to write rule-like prompts.
Building on top of our findings, we envision a hybrid approach to custom classifier creation: users could label examples to bootstrap and iterate on their prompt classifiers, while decomposing a complex preference into a rule structure could provide users with greater transparency.


\begin{acks}
This research was supported by NSF award \#2236618. We would like to thank members of the Social Futures Lab at the University of Washington for their invaluable help in this project. We also would like to thank our anonymous reviewers for their insightful feedback. Finally, we would like to express our heartfelt thanks to all the participants who dedicated their time and effort to participate in our study.
\end{acks}


\bibliographystyle{ACM-Reference-Format}
\bibliography{filterbuddy}

\appendix
\label{appendix}

\section{Pilot Studies}

We conducted pilot studies with our lab members and 5 non-technical participants to iteratively implement our three systems. We included non-programmers in these studies to guarantee that our interfaces would be user-friendly and accessible to all potential users.

\subsection{\textsc{Label System}}

Since both the \textsc{Rule System} and the \textsc{Prompt System} allowed users to review the performance of their filters during the creating process, we tested real-time feedback for the \textsc{Label System} in our pilot studies. Specifically, when users were labeling examples from the new batch of active learning, they could review how the filter that was trained on previous labels would predict on each example. We experimented with displaying this information immediately after users labeled an example and after they labeled all examples in a batch. However,  our pilot studies suggested that such features were more confusing than helpful for end-users, so we excluded them from the \textsc{Label System} for our experiments.

\subsection{\textsc{Rule System}}

We used our pilot studies to iteratively develop a more user-friendly \textsc{Rule System} than AutoMod that is tailored to the moderation needs of individual users. Specifically, we made the following changes to adapt the AutoMod system for personal moderation.

\begin{itemize}[leftmargin=0.5cm]
    \item AutoMod permits a variety of conditions in a rule such as ``include,'' ``exclude,'' ``start with,'' or even regular expressions. However, regular expressions might be excessively difficult for end-users, and certain conditions, like ``start with,'' may not be as beneficial for filtering unwanted content. To simplify the rule creation process, our system restricts end-users to using only ``include'' and ``exclude'' conditions. We call the exclude condition an ``exception'' because users might be confused by a complicated rule like ``remove a unit that includes this word but excludes that word.''

    \item While AutoMod allows an unlimited number of conditions per rule, this complexity can be overwhelming and not particularly useful for end-users in expressing their preferences. Therefore, we limit users to having at most two ``include'' conditions and one ``exclude'' condition per rule to maintain simplicity and clarity.
    
    \item AutoMod offers a range of actions for each rule, including approving and removing caught texts. In contrast, our system is designed solely for creating rules that remove texts. This design choice is based on observations that navigating between ``approve''/``remove'' actions and ``include''/``exclude'' conditions can be confusing for end-users.

    \item Prior research has indicated that users often struggle to account for all potential spelling variants that could bypass word filters~\cite{jhaver2022designing}. To reduce the mental load of brainstorming synonyms for phrases, we integrated a ``similar phrases'' feature that leverages the LLM to suggest phrases that are similar to existing ones provided by users. Additionally, our system offers an option to detect the following spelling variants. We implemented the detection of the first three types of variants by generating generalized regular expressions for each phrase. For nouns and verbs, we first identify the part-of-speech tag of the word~\cite{chen2014fast}, then search for its plural form if it is a noun or its various tenses if it is a verb~\cite{de2012pattern}.
    
     \begin{itemize}
         \item phrases with repeated letters, such as ``coooool'' for ``cool''

         \item phrases with mixed uppercase and lowercase letters, such as ``Cool'' for ``cool''

         \item phrases where letters are replaced with visually similar characters, such as ``co0l'' for ``cool''

         \item singular or plural forms of nouns, such as ``apples'' for ``apple''

         \item different tenses of verbs, such as ``found'' for ``find''
     \end{itemize}
\end{itemize}

\subsection{\textsc{Prompt System}}

We evaluated two potential methods for the \textsc{Prompt System}. First, we put all user-generated prompts into a single system prompt template and only queried the LLM once for the overall prediction. This approach compromised the quality of LLM predictions since the LLM sometimes skipped a few examples or completely ignored certain prompts. 
Second, we queried the LLM multiple times for each prompt and determined the final prediction based on the majority. This quality-minded approach often required users to wait more than 5 minutes for LLM predictions and thus discouraged users from interactively testing their filters during the creation process. 

Some may advise prioritizing the quality of LLM predictions during the filter creation process in order to evaluate the full potential of LLMs in personal moderation. However, content moderation algorithms in deployment often need to analyze vast quantities of comments per day. Prioritizing quality without regard for efficiency can therefore lead to impractical and costly operations. To ensure that users accurately evaluate how well their filters would work once deployed, it is essential that, during the filter creation process, users engage with the algorithm under conditions that mirror actual deployment settings. Thus, we opted to prioritize a low response time while ensuring that prediction quality was still sufficient in order to mirror realistic deployment settings.
Ultimately, we used the following system prompt:

\texttt{For each text in the dataset, give a 1 (True) or 0 (False) prediction to represent whether the text satisfies the description in the rubrics. Each text starts with “DATA” and a number. Both the number and the text are enclosed by “<” and “>”.}

\texttt{In the following, the user will provide one rubric to help you make your decision. If the given rubric is completely satisfied, give a True prediction. Otherwise, give a False prediction.}

\texttt{RETURN YOUR ANSWER in the JSON format \{\{“results”: [(index, prediction), ...]\}\} where (index, prediction) is a tuple, the index is the number of the text in the dataset, and prediction is either 1 or 0.}

\leijie{\section{Examples of User-Created Rules/Prompts}}

\leijie{Here we list three examples of prompts that participants authored in our experiments.}

\begin{itemize}
    \item \textit{Remove texts that refer to people as stupid, dumb, idiots, evil, useless, mentally ill, inbreds, nuts, idiot, demented, deranged, scum, retard.}

    \item \textit{Remove off-topic texts such as self-promotion unrelated to the current conversation.}

    \item \textit{Delete texts that disclose private, sensitive personal information (e.g., addresses, phone numbers) unrelated to the discussed topic.}
\end{itemize}

\leijie{In addition, we list three examples of rules that participants authored in our experiments.}

\begin{itemize}
    \item Remove comments that include at least one of the following phrases \textit{``bitch, stupid, crazy, bitches, stupidity, mad, craziness, dumb, asshole, lunatic, idiot, dick, crackpot, psycho, dumbass, idiots, nutcase, psychos, assholes, moron, freak''}.

    \item Remove comments that include at least one of the following phrases \textit{``republican, Republicans, GOP, conservative, right-wing, Grand Old Party, democrats, conservatives, right-wingers, Republican Party, GOP members, Democratic Party''} and at least one of the following phrases \textit{``kill, dead, violence''}.

    \item Remove comments that include at least one of the following phrases \textit{``keep them away from African Americans, keep away from blacks, stay away from African Americans, avoid African Americans, keep them from blacks, Republicans are idiots, You bonehead Democrats, Avoid blacks, Stay away from blacks, Keep away from African American, Stay clear of African Americans''}.

\end{itemize}


\end{document}